\documentclass[12pt,preprint]{aastex}

\newcommand\simgt{\lower.5ex\hbox{$\; \buildrel > \over \sim \;$}}
\newcommand\simlt{\lower.5ex\hbox{$\; \buildrel < \over \sim \;$}}
                                                                                
\newcommand\cs{c_s}
\newcommand\vA{v_{\rm A}}
\newcommand\Msun{\;M_\odot}
\newcommand\kms{\rm\;km\;s^{-1}}
\newcommand\Sden{\;M_\odot{\rm\;pc^{-2}}}
\newcommand\Ssp{\Sigma_{\rm sp}}
\newcommand\torb{t_{\rm orb}}
\newcommand\tgrow{t_{\rm grow}}
\newcommand\pc{{\rm pc}}
                                                                                
\newcommand\sles{\lower2pt \hbox{$\;\buildrel {\scriptstyle <}
   \over {\scriptstyle\sim}\;$}}
\newcommand\sgreat{\lower2pt\hbox{$\;\buildrel {\scriptstyle >}
   \over {\scriptstyle\sim}\;$}}
                                                                                
\slugcomment{Accepted for publication in ApJ}
\shorttitle{3D Spur and Cloud Formation}
\shortauthors{Kim \& Ostriker}

\begin{document}

\title{Formation of Spiral-Arm Spurs and Bound Clouds in Vertically
Stratified Galactic Gas Disks}

\author{Woong-Tae Kim}
\affil{Astronomy Program, SEES, Seoul National University,
Seoul 151-742, Republic of Korea}
\email{wkim@astro.snu.ac.kr}
\and
\author{Eve C. Ostriker}
\affil{Department of Astronomy, University of Maryland \\
College Park, MD 20742, USA}
\email{ostriker@astro.umd.edu}
                                                                                
\begin{abstract}

We investigate the growth of
spiral-arm substructure in vertically stratified, self-gravitating, 
galactic gas disks, using local numerical MHD simulations.  Our new
models extend our previous two-dimensional studies \citep{kim02a},
which showed that a magnetized spiral shock in a thin 
disk can undergo magneto-Jeans instability (MJI), resulting in
regularly-spaced interarm spur
structures and massive gravitationally-bound fragments.  Similar spur
(or ``feather'') features have recently been seen in high-resolution 
observations of
several galaxies, and massive bound gas condensations are likely the 
precursors of giant molecular cloud complexes (GMCs) and \ion{H}{2} regions.
Here, we consider two sets of numerical models: two-dimensional
simulations that use a ``thick-disk'' gravitational kernel, and 
three-dimensional simulations with explicit vertical stratification.  
Both models adopt an isothermal equation of state with $\cs=7 \kms$.
When disks are sufficiently magnetized and self-gravitating, the
result in both sorts of models is the growth of 
spiral-arm substructure similar to that in our previous razor-thin models.
Reduced self-gravity due to nonzero disk thickness increases 
the spur spacing to $\sim 10$ times the Jeans length at the arm
peak, a factor $\sim3-5$ times larger than in razor-thin models.
Bound clouds that form from spur fragmentation have masses 
$\sim (1-3)\times  10^7\Msun$ each, a factor $\sim3-8$ times larger than
in razor-thin models 
with the same gas surface density and stellar spiral arm strength.
These condensation masses are comparable to results from
other three-dimensional models without spiral structure, and similar
to the largest observed GMCs.  
The mass-to-flux ratios and specific angular momenta of the bound
condensations are lower than large-scale galactic values, as is true for
observed GMCs.
We find that unmagnetized or weakly magnetized
two-dimensional models are unstable to the ``wiggle instability'' 
previously identified by \citet{wad04}, and proposed as a potential
spur- and clump-forming mechanism.  However, 
our fully three-dimensional
models do not show this effect.   
Non-steady motions and strong vertical shear prevent coherent vortical
structures from forming, evidently suppressing the wiggle instability
that appears in two-dimensional (isothermal) simulations.  
We also find no clear traces of Parker instability in the nonlinear 
spiral arm substructures that emerge (in self-gravitating models),
although conceivably Parker modes may help seed the MJI at early stages since
azimuthal wavelengths are similar.
\end{abstract}
                                                                                

\section{Introduction}

The interstellar medium (ISM) contains structure at many scales, and
the growth of much of this structure is believed to represent the
first step in initiating galactic star formation.  In particular,
the molecular portion of the ISM is concentrated into dense clouds
over a range of scales \citep{bli93}; each of these clouds also contains
significant higher-density substructure.  The largest of these cloud
complexes (up to $\sim 10^7\Msun$, including atomic envelopes) 
contain most of the mass in the overall distribution of
giant molecular clouds (GMCs).  In galaxies with prominent spiral
structure, GMCs tend to be concentrated into the spiral arms.
Observationally, the higher-than-average specific star formation rates
(i.e.\ per unit gas mass) within spiral arms \citep{kna92,kna96} suggests an
evolutionary sequence: compression of diffuse gas (and/or collection
of smaller clouds) within the arms prompts GMC formation, and star formation
is subsequently triggered in the densest portions within GMCs.

From a theoretical point of view, interaction of the ISM 
with spiral arms should
naturally result in bound cloud formation when the mean density is
high enough that self-gravitating instabilities are able to grow.  The
growth of self-gravitating structures in spiral arms has been
investigated by many authors, using both linear-theory analyses 
\citep{bal85,bal88,elm94,kim02a}
and more recently, nonlinear hydrodynamic and magnetohydrodynamic
(MHD) simulations that yield bound clouds with properties similar to
observed GMCs (\citealt{kim02a}, hereafter Paper I).

Interestingly, self-gravity acting on the gas within spiral arms can
apparently also  lead to growth of structures larger than GMCs.  These
structures take the shape of gaseous spurs that project from the main
body of the arm into the interarm region.  The first direct numerical
simulations showing the development of these trailing spurs
(Paper I) occurred contemporaneously with
the release of the now-famous {\it Hubble Heritage} image of M51
\citep{scorec01,sco01}, in which strikingly similar structures are
evident.  Furthermore, a recent {\it Spitzer Legacy} image of M51
displays a quasi-regular distribution of thin, trailing dust filaments
throughout the interarm regions \citep{ken04}.  The dust filaments
seen in these images could well be the remnants of gaseous spurs
initiated inside spiral arms, which have maintained their integrity
deep into the interarm regions.  A recent archival study of HST galaxy
images (Lavigne, Vogel, \& Ostriker 2006, in preparation) 
has shown that arm substructures of the kind seen
in M51 are in fact ubiquitous in the grand design spirals where the
global pattern is clean enough to make identification of spiral spur
substructures possible.  When identified as a regular series of dust
lanes extending out of a primary dust lane across the bright stellar
arm and into the interarm region,
such spur structures have historically been termed ``feathers''
\citep{lyn70}.

In this work, we extend the (thin disk) two-dimensional numerical
models of Paper I into three dimensions, in order to study the effects
of vertical stratification on the development of condensation
instabilities in spiral arms.  In our previous work, we argued that
self-gravity, aided by magnetic fields, is the key to gaseous spur and
dense cloud formation.  However, it has recently been proposed that
Kelvin-Helmholtz instabilities are also able to trigger spiral arm
spur formation \citep{wad04}.  In addition, a theoretical proposal
with a long history is that Parker instabilities in spiral arms may be
important in prompting GMC formation (e.g.
\citealt{par66,mou_etal74,bli80}; see discussion and references in
\citealt{kim02b}, hereafter Paper II).  Our three-dimensional models
allow us to explore both of these proposals using more realistic
numerical set-ups than in previous work.

We shall show that, for vertically stratified magnetized disks,
qualitatively quite similar results to the conclusions of Paper I
appear to hold.  Quantitatively, we shall also show that the
three-dimensional results are consistent with results of
two-dimensional models when ``thick-disk'' dilution of gravity is
incorporated.  We also find that three-dimensional dynamics tends to
suppress the growth of Kelvin-Helmholtz-driven spurs described by
\cite{wad04} based on unmagnetized two-dimensional simulations, and 
that there are
no clear signatures of Parker instability in our models.

Star formation takes place within molecular clouds, and at least the
most massive GMCs that dominate the distribution appear to live for
significantly less than a galactic orbital time, based on their close
association with spiral arms \citep{sol85,sol89,eng03,sta05b}.
Thus, the galactic star formation rate $\dot M_*$ at a given radius
can be written as a cloud formation rate $\dot M_{\rm cloud}$ times a star
formation efficiency per cloud lifetime, $\epsilon_{\rm SF}$.  To the
extent that this star formation efficiency depends primarily on
physical processes within a cloud, understanding the regulation of
star formation can therefore be separated into a large-scale (galaxy)
and a small-scale (GMC) problem.  The current work primarily addresses
the large-scale problem by investigating the mechanisms and timescales 
required for
diffuse gas to be gathered into bound clouds in the case when strong
spiral structure is present.  It also informs the small-scale problem
by ascertaining the properties (in terms of masses, magnetic fluxes,
and spins) of the bound clouds that form, which affect the star
formation efficiency.  A complete theory of cloud formation should
start from initial conditions that would be left by the previous
generation of clouds, in which feedback from star formation has taken
its toll.  Clouds may be partially destroyed by photoevaporation and
partly disaggregated by stellar winds and supernovae.  We defer this
more complex treatment for the future, here taking the first step of
considering cloud formation in a medium that is initially 
relatively uniform in its large-scale properties.

The plan of this paper is as follows:  In \S 2, we lay out our
numerical methods and the specifications of the models we shall study.
In \S 3, we present results for two-dimensional ``thick disk gravity''
models, in which both magneto-Jeans instability (see Paper I) and the
``wiggle instability'' of \cite{wad04} can develop, depending on
parameters.  In \S 4, we present the results of three-dimensional
stratified models, identifying the conditions under which spurs and clouds
form via various different self-gravitating mechanisms (but not
through the ``wiggle instability'').  In \S 5, we conclude with a
summary of our new results, and a discussion of their implications
within the larger context of current work on galactic structure and
star formation.

\section{Numerical Methods and Model Parameters}

In this paper, we study the nonlinear evolution of vertically stratified, 
differentially rotating, self-gravitating, magnetized galactic gas disks under 
spiral stellar potentials using local, three-dimensional  
numerical simulations.  Similar work in a two-dimensional, razor-thin 
disk geometry was reported in Paper I, while three-dimensional
evolution of gas 
in the absence of the spiral potential perturbation was studied
in Paper II.  In this section we briefly summarize the numerical methods 
and model parameters that we adopt; the reader is refereed to Papers I and II 
for the more complete description.

\subsection{Basic Equations and Numerical Methods}

We study the nonlinear evolution of self-gravitating gas under the influence of
a stellar 
spiral potential that is tightly wound with a pitch
angle $i\ll1$, and rigidly rotating at a constant pattern speed  $\Omega_p$
with respect to an inertial frame.
For local simulations (i.e. domain $\ll R_0$)
involving spiral arms, it is advantageous to 
construct a local Cartesian frame centered on a position 
$R_0$, $\phi_0=\Omega_p t$, $z_0=0$, that 
corotates with the stellar spiral pattern.  The local Cartesian frame 
is inclined with respect to the $\hat R$, $\hat \phi$ coordinate directions
by an angle $i$ in such a way that 
$\bf \hat{x}$ and $\bf \hat{y}$ correspond to the directions in the midplane
perpendicular and parallel, respectively,
to the local segment of the spiral arm, while $\bf \hat{z}$ denotes
the direction perpendicular to the galactic plane \citep{rob69,bal88}.
The simulation domain is a rectangular parallelepiped with size 
$L_x \times L_y \times L_z$.
The dimensions of the box are $L_x=2\pi R_0\sin i/m$, equal to
the arm-to-arm distance for an $m$-armed spiral,  
$L_y=2L_x$, and  $L_z=8H_0$, where
$H_0$ denotes the vertical scale height of the gas distribution 
when the spiral potential perturbation is absent. 

In this local frame, the background velocity arising purely 
from galactic rotation
is approximately given by 
\begin{equation}\label{v0}
{\bf v}_0= R_0(\Omega_0-\Omega_p)\sin i {\bf\hat x}
+ [R_0 (\Omega_0-\Omega_p) - q_0 \Omega_0x] {\bf\hat y},
\end{equation}
where $\Omega_0$ is the angular velocity of gas at $R_0$ in the 
inertial frame 
and $q_0\equiv -(d\ln\Omega /d\ln R)|_{R_0}$ is the local shear
parameter of the background flow in the absence of the spiral arm (Paper I).
For a flat rotation curve, $q_0=1$. 
We assume that the gas velocity $\bf v$ induced by the spiral potential 
is much smaller than $R_0\Omega_0$, 
and neglect terms arising from curvature effects in the coordinates 
(e.g., \citealt{gol65b,jul66}).
In this local, rotating frame, the MHD equations are written as
\begin{equation}\label{con}
  \frac{\partial\rho}{\partial t} + \nabla\cdot(\rho \mathbf{v}_T) = 0,
\end{equation}
\begin{equation}\label{mom}
  \frac{\partial\mathbf{v}}{\partial t} +
        \mathbf{v}_T\cdot\nabla\mathbf{v}
     =  -\frac{1}{\rho}\nabla P
       + \frac{1}{4\pi\rho}(\nabla\times\mathbf{B})\times\mathbf{B} 
        + q_0\Omega_0 v_x \mathbf{\hat{y}}
       - 2\mathbf{\Omega}_0\times\mathbf{v}
       - \nabla (\Phi_{\rm s} + \Phi_{\rm ext}),
\end{equation}
\begin{equation}\label{ind}
  \frac{\partial \mathbf{B}}{\partial t} =
       \nabla\times(\mathbf{v}_T\times\mathbf{B}),
\end{equation}
\begin{equation}\label{Pos}
   \nabla^2\Phi_{\rm s} = 4\pi G \rho,
\end{equation}
and
\begin{equation}\label{eos}
P = c_s^2\, \rho,
\end{equation}
(cf.\ \citealt{rob69,rob70,shu73,bal85,bal88}),
where ${\mathbf v_T} \equiv {\mathbf v_0} + {\mathbf v}$ is the total 
velocity in the rotating frame, $c_s$ is the isothermal sound speed, 
$\Phi_{\rm s}$ is the self-gravitational potential of the gas,
and $\Phi_{\rm ext}$  is the external stellar potential.
Other symbols have their usual meanings.
In all the simulations presented in this paper, for simplicity we adopt
an isothermal equation of state in both space and time,
as expressed by equation (\ref{eos}).

The imposed external stellar potential $\Phi_{\rm ext}$ varies both in the
plane and perpendicular to the plane; to lowest order it is separable into 
two parts as
\begin{equation}\label{ext_pot}
\Phi_{\rm ext} = 
\left(\frac{\pi G \Sigma_*}{\sigma_{*,z}}\right)^2z^2 + 
\Phi_{\rm sp} \cos \left(\frac{2\pi x}{L_x}\right).
\end{equation}
The first, quadratic term in $z$ describes the variation in
the gravitational potential near the midplane from a self-gravitating stellar 
distribution with surface density $\Sigma_*$ and vertical velocity dispersion
$\sigma_{*,z}$.\footnote{The 
formula for the gravitational
potential of a self-gravitating stellar disk treated as an isothermal fluid 
is 
$\Phi_*(z)
= 2\sigma_{*,z}^2\ln \cosh({z/H_*})$, where $H_*\equiv 
\sigma_{*,z}^2/(\pi G \Sigma_*)$ (cf.\ \citealt{gil90}). The central
stellar
density is $\rho_{0,*}=\Sigma_*/(2H_*)$.
For
$z/H_*\ll1$, $\Phi_*(z)\approx  (\sigma_{*,z}/H_*)^2 z^2=
2\pi G \rho_{0,*} z^2= (\pi G \Sigma_*/\sigma_{*,z})^2 z^2.$
}

Since the bulk of gas remains within one scale height of the stellar disk,
this is quite a good approximation for the stellar potential in
studying the dynamical evolution of the gas. 
The second, sinusoidal term with amplitude $\Phi_{\rm sp}$ in 
equation (\ref{ext_pot}) represents a local
analog of the logarithmic spiral 
potential used in \citet{rob69} and \citet{shu73}.
Since $\Phi_{\rm sp}<0$ and 
$|x|\leq L_x/2$,  the spiral potential attains its minimum at 
the center ($x=0$) of the box. 

We integrate the time-dependent, ideal MHD equations 
(\ref{con})-(\ref{ext_pot}) using a modified version of the ZEUS code
\citep{sto92a,sto92b}.  It uses a time-explicit, operator-split,
finite-difference scheme for solving the MHD equations on a staggered
mesh, and employs the ``constrained transport'' and ``method of 
characteristics'' algorithms to maintain the divergence free condition
of magnetic fields as well as to ensure accurate propagation of 
Alfv\'en waves.  For less diffusive transport of hydrodynamic variables,
we apply a velocity decomposition method in updating $v_y$ \citep{kim01}.
We also
employ the Alfv\'en limiter algorithm of \citet{mil00}, setting
the limiting speed of the displacement current to $c_{\rm lim}=8\cs$; 
this value of $c_{\rm lim}$ allows a good dynamical range in the 
low-density, high-$z$ regions.
We solve the Poisson equation by combining the fast Fourier transform
method in sheared horizontal coordinates \citep{gam01}
with the Green function method for vertical integration \citep{miy87}.
We adopt the shearing box boundary conditions of \citet{haw95}
in which the $x$-boundaries are shearing-periodic and
the  $y$-boundaries are perfectly periodic,
while implementing the outflow boundary conditions of 
\citet{sto92a} at the $z$-boundaries.

We have parallelized the numerical code we use with both Open-MP and
MPI, for use on both shared- and  distributed-memory platforms.  The
parallelization is achieved by domain decomposition
along the $z$-direction, which is convenient for our hybrid technique of
solving the Poisson equation.  Our high resolution models in three
dimensions have $128 \times 256\times 128$ zones in $(x,y,z)$.

\subsection{Model Parameters}

The ISM in galaxies is highly inhomogeneous, turbulent, and multi-phase
(e.g., \citealt{fie69,mck77,hei01,heil05,wol03}),
and a fully realistic treatment of ISM evolution entails consideration of 
heating, cooling, and other physical processes that may significantly 
affect the density and temperature structures of the gas 
(e.g, \citealt{vaz00,kri04,pio04,aud05,pio05}). 
In order to focus on the role of self-gravity in forming intermediate-scale 
spiral-arm
substructure, we instead consider for all models presented in this paper 
homogeneous initial gas distributions, adopting a simple isothermal
equation of state with an ``effective'' speed of sound $\cs$.
The effects of ISM heating/cooling and the associated multi-phase cloudy 
gas distribution on the interaction with 
spiral arms will be considered in a subsequent paper.

Our initial disks, in the absence of the spiral perturbation 
($\Phi_{\rm sp}=0$ in eq.\ [\ref{ext_pot}]), are vertically
stratified with density $\rho_0(z)$, have a uniform surface density
$\Sigma_0=\int\rho_0(z)dz$, and are 
threaded by magnetic fields ${\bf B_0}=B_0(z){\bf\hat y}$ pointing
parallel to the spiral arm.
The initial equilibrium satisfies force balance between the 
total (thermal plus magnetic) pressure gradient and the total 
(self plus external) gravity along the $z$-direction.  
For the strength of the external
vertical gravity, we define $s_0 \equiv (\sigma_{*,z}\Sigma_0)^2/
(\cs\Sigma_*)^2$ and take $s_0=1$.  This choice of 
$s_0$ corresponds to 
average galactic-disk conditions of $\sigma_{*,z}\sim 20\kms$,
$\Sigma_*\approx 35\Sden$, and total gas surface density 
$\Sigma_0\approx (11-16)\Sden$
(e.g., \citealt{kui89,hol00}).
Physically, $s_0$ measures the ratio of gaseous self-gravity to stellar gravity
at one scale height $H_0$ of the gas;
$H_0=\Sigma_0/(2\rho_0[0])\approx 200$ 
pc for the adopted sets of parameters in this paper (see below and 
Paper II for a 
discussion of the vertical equilibrium specification).

The three key dimensionless 
parameters that characterize our simulation models are
\begin{equation}\label{Q0}
Q_0 \equiv \frac{\kappa_0\cs}{\pi G \Sigma_0}
= 1.4 \left(\frac{\cs}{7.0 {\rm\,km\,s^{-1}}} \right)
      \left(\frac{\kappa_0}{36\,{\rm km\,s^{-1}\,kpc^{-1}}}\right)
      \left(\frac{\Sigma_0}{13\,\Msun\pc^{-2}}\right)^{-1},
\end{equation}
\begin{equation}\label{beta0}
\beta_0 \equiv \frac{\cs^2}{\vA^2} = \frac{4\pi\rho_0(z)\cs^2}{B_0^2(z)}, 
\end{equation}
and
\begin{equation}\label{F0}
F \equiv \frac{m}{\sin i}
\left(\frac{|\Phi_{\rm sp}|}{R_0^2\Omega_0^2}\right),
\end{equation}
where $\vA= B/(4\pi\rho)^{1/2}$ is the Alfv\'en speed.
The Toomre stability parameter $Q_0$ describes the gas surface density 
relative to various critical values that mark thresholds for 
{\it axisymmetric} gravitational 
instability.  For a razor-thin disk, the instability threshold is 
at $Q_c=1$ \citep{too64}; for  a purely self-gravitating
($s_0=\infty$), vertically-stratified  disk, 
the threshold is at $Q_c\approx0.676$  
\citep{gol65a,gam01}; and for a disk 
subject to both self-gravity and external gravity with $s_0=1$, the threshold 
is at $Q_c\approx0.75$ (Paper II; \citealt{kim03}, hereafter Paper III).  
The dimensionless parameter $F$ in equation (\ref{F0})
measures the ratio of the maximum perturbation force from the external 
spiral potential to the  
mean radial gravitational force \citep{rob69}.
Initially, we take $\beta_0$ to be constant everywhere, so that
$B_0(z)\propto [\rho_0(z)]^{1/2}$.

To represent solar neighborhood conditions in dimensional quantities, we use
galactocentric radius $R_0=10$ kpc.  With 
$\Omega_0=26\kms\;{\rm kpc^{-1}}$, the local epicyclic frequency is
$\kappa_0\equiv (R^{-3}d[R^4\Omega^2]/dR)|_{R_0})^{1/2} 
=(4-2q_0)^{1/2}\Omega_0\approx 36\kms\;{\rm kpc^{-1}}$
for a nearly flat rotation curve with $q_0\approx 1$ \citep{bin87}.
The corresponding galactic orbital period is 
$\torb\equiv 2\pi/\Omega_0
=2.4\times 10^8\;{\rm yr}\;(\Omega_0/26\kms\;{\rm kpc^{-1}})^{-1}$,
which we adopt as a fiducial time unit in our presentation.
Our results can be rescaled to use other galactic parameters, provided that
the dimensionless ratios $\Omega_p/\Omega_0$, $L_x/R_0$, and 
$\Omega_0 R_0/c_s$  (as well as 
$q_0$, $Q_0$, $\beta_0$, and $F$) remain the same.

For spiral arm parameters, we adopt for all our models pattern speed
$\Omega_p=\Omega_0/2$, pitch angle $\sin i =0.1$, and azimuthal
wavenumber $m=2$; the corresponding arm-to-arm distance is $L_x=2\pi
R_0\sin i/m$.  For our fiducial parameters, the quasi-radial
size of the simulation domain is therefore $L_x=3.14$ kpc.

To simulate various galactic conditions, we select
the following three sets of the parameters:
($Q_0$, $\beta_0$, $F)=(1.8, \infty$, 5\%),
(1.5, 10, 5\%), and (1.2, 1, 10\%).  For our fiducial $R_0$ and $\Omega_0$, 
and taking $c_s= 7 \kms$,
the corresponding gas surface density,
total gas mass $M_{\rm tot}$ contained in the simulation box, 
and scale height $H_0$ are
11$\Sden$, $2.3\times10^8\Msun$, and $200$ pc for the $Q_0=1.8$ model
with $\beta_0=\infty$;
13$\Sden$, $2.8\times10^8\Msun$, and $170$ pc for the $Q_0=1.5$ model with
$\beta_0=10$; and
16$\Sden$, $3.4\times10^8\Msun$, and $180$ pc for the $Q_0=1.2$ model 
with $\beta_0=1$.  For the $\beta_0=1$ and 10 models, the respective midplane
magnetic field strengths are $4.4\mu$G and $1.3\mu$G, respectively.
The model parameter sets are chosen
to ensure that all the models are gravitationally stable to 
{\it quasi-axisymmetric} (wavefronts parallel to the spiral arm) perturbations 
even when the effect of the spiral potential is included.
Our objective in this paper is to explore how 
{\it nonaxisymmetric} disturbances in these 
systems evolve, subject to interactions with spiral arms.  We also
briefly consider, in \S 4.2.4, development of a model which is
unstable to quasi-axisymmetric disturbances.

\section{Two-Dimensional Models with Thick-Disk Gravity} 

Paper I studied the formation of spiral-arm and interarm substructure
in two-dimensional disks that were treated as being infinitesimally thin.  
One of the drawbacks of the thin-disk approximation of Paper I is that it
overestimates self-gravity at the disk midplane for modes whose
wavelengths approach the disk scale height (e.g.,
\citealt{too64,elm87}).  For a given strength of the spiral arm
potential, therefore, razor-thin disks tend to have larger density
enhancement at the spiral shock than vertically-resolved disks,
increasing the susceptibility to gravitational instability.
Consequently, the critical $F$ values for stable, one-dimensional
spiral shock configurations are lower in razor-thin disks than in more
realistic vertically extended disks; to ensure gravitational stability to
quasi-axisymmetric perturbations, razor-thin disk models considered in
Paper I were limited to $F\leq 3\%$ for a range of $Q_0$ and $\beta_0$
values.

We revisit two-dimensional disk models here, now 
taking into account the geometrical dilution 
of self-gravity due to finite disk thickness.  
Instead of the standard kernel appropriate for a thin disk, we use
a ``thick-disk'' gravitational kernel such that the
potential and density Fourier modes with wavenumber $k$ at the midplane
are related by
\begin{equation}\label{kernel}
\Phi_{\rm s}(k) = - \frac{4\pi GH_0\rho(k)}{|k|(1+|k|H_0)},
\;\;\;{\rm for}\;k\neq0,
\end{equation}
and $\Phi_{\rm s}(k)=0$ for $k=0$, 
where the disk scale height $H_0$ is held fixed in both space and time.
Equation (\ref{kernel}) is in fact exact 
for an exponential density distribution $\rho\propto e^{-z/H_0}$
(e.g., \citealt{elm87}), and generally yields a result within 15\% of the 
exact solution of the three-dimensional Poisson equation
for self-consistent vertical equilibria (Paper II). 
For our two-dimensional models, the dependences of all fluid variables on 
the vertical coordinate are neglected.
The results of these two-dimensional disk models allow us to quantify the impact 
of thick-disk gravity on the formation of arm/interarm substructure
without the potential influence of dynamical instabilities
that critically involve the vertical dimension.

We have run three sets of models with differing numerical resolutions. 
The model parameters and simulation results are given
in Table \ref{tbl-1}.  Column (1) labels each run.
Columns (2) and (3) list the Toomre $Q_0$ parameter (eq.\ [\ref{Q0}]) and
the field strength in terms of the plasma parameter 
$\beta_0$ (eq.\ [\ref{beta0}]) of an initial disk when a spiral arm
perturbation is absent, while column (4) gives the amplitude of 
the spiral potential in terms of $F$ (eq.\ [\ref{F0}]).
Column (5) indicates the numerical resolution $N_x \times N_y$ 
of the model.
Since our simulation box has $L_y=2L_x$, $N_y=2N_x$ makes each cell
square in the $x$-$y$ plane.
The peak surface density $\Sigma_{\rm sp}$ of the resulting spiral shock
configuration is given in column (6), while column (7) gives the 
corresponding local value 
$Q_{\rm sp}=Q_0(\Sigma_{\rm sp}/\Sigma_0)^{-1/2}$ at the spiral density peak; 
this scaling follows from the constraint of potential
vorticity conservation (\citealt{hun64,bal88,gam96}; Paper I).
The width $W$ of the gaseous spiral arm defined at
$\Sigma = ({\rm max}[\Sigma]+{\rm min}[\Sigma])/2$ is listed in column (8).  
Finally, the mean separation ($\lambda_y$) of the structures that develop 
in each model is given in column (9) in terms of the arm-to-arm 
distance $L_x$ and in column (10) in terms of the normalized wavenumber
$K_{y} \equiv \lambda_{\rm J,sp}/\lambda_y$, where the local, thin-disk
Jeans wavelength at the arm density peak is defined by
\begin{equation}\label{lambda}
\lambda_{\rm J,sp} = \frac{\cs^2}{G\Sigma_{\rm sp}} = 160 \pc 
\left({\cs \over 7 \kms}  \right)^2 
\left(\Sigma_{\rm sp}\over 70\Sden \right)^{-1}.
\end{equation}
Note that structures that form from magneto-Jeans instability (MJI) in a 
featureless, uniform, low-shear,  razor-thin disk favor
$K_{y}=0.50-0.75$ \citep{kim01}, 
while MJI-driven fragmentation
occurring inside spiral arms in the razor-thin limit 
was found to have $K_{y}\sim0.45-0.54$ (Paper I). 

To initiate our models, we begin with a disk having 
a uniform shear profile expressed by equation (\ref{v0}), and
uniform surface density and magnetic fields corresponding to the
chosen values of $Q_0$ and $\beta_0$.  Using a one-dimensional grid in 
$\hat x$, we turn on the 
external spiral potential and slowly increase its amplitude 
up to a desired level, $F$.  This yields a one-dimensional 
equilibrium spiral shock 
profile, which we then use to initialize our two-dimensional simulations.
On top of the background profile,
we apply density perturbations created by 
a Gaussian random field with flat power 
for $1\leq kL_x/2\pi\leq 128$ and zero power for $kL_x/2\pi>128$.
The standard deviation of the density perturbations is fixed to
be 3\% in real space.\footnote{This flat power spectrum with low amplitudes 
by no means represents realistic interstellar perturbations; it is chosen 
simply to allow identification and study 
the linear and nonlinear responses of the most dominant modes of the
system.  The ultimate outcomes of the simulations are not, however,
sensitive to this choice.}
We then follow the two-dimensional flow as
perturbations evolve and grow through interaction with the 
background spiral shock, eventually forming self-gravitating substructures.
Note that an alternative initialization procedure is simply to start with 
a uniformly-shearing disk, and slowly turn on the spiral forcing term; this is
the procedure we follow for our three-dimensional models (see \S 4),
and we have also tested two-dimensional models with this method.  
We find, for our
two-dimensional models, essentially the same end results regardless of the 
initialization procedure. 

\subsection{Magneto-Jeans Instability inside Spiral Arms} 

In this subsection we describe the evolution of model ME2d1 (and its
lower-resolution counterparts ME2d2-3),
which have equipartition magnetic fields.  In
these simulations, gaseous spurs and gravitationally bound clouds form
as a direct consequence of MJI operating within spiral arms.  
The phrase `bound clouds' in this paper refers to gaseous clumps that 
would collapse in a runaway fashion if physical processes such 
as turbulence, fragmentation, or star formation did not subsequently limit 
this collapse.
We note
that if the razor-thin gravitational kernel had been used (i.e.
$H_0=0$ in eq.\ [\ref{kernel}]), then the spiral shocks produced by
this set of model parameters ($Q_0=1.2$, $\beta_0=1$, $F=10\%$) would
have been gravitationally unstable to quasi-axisymmetric
perturbations.  Instead, with nonzero disk thickness, a
quasi-axisymmetric, stable shock configuration is possible at a value
of $F=10\%$.

\begin{figure}
\epsscale{1.}
\plotone{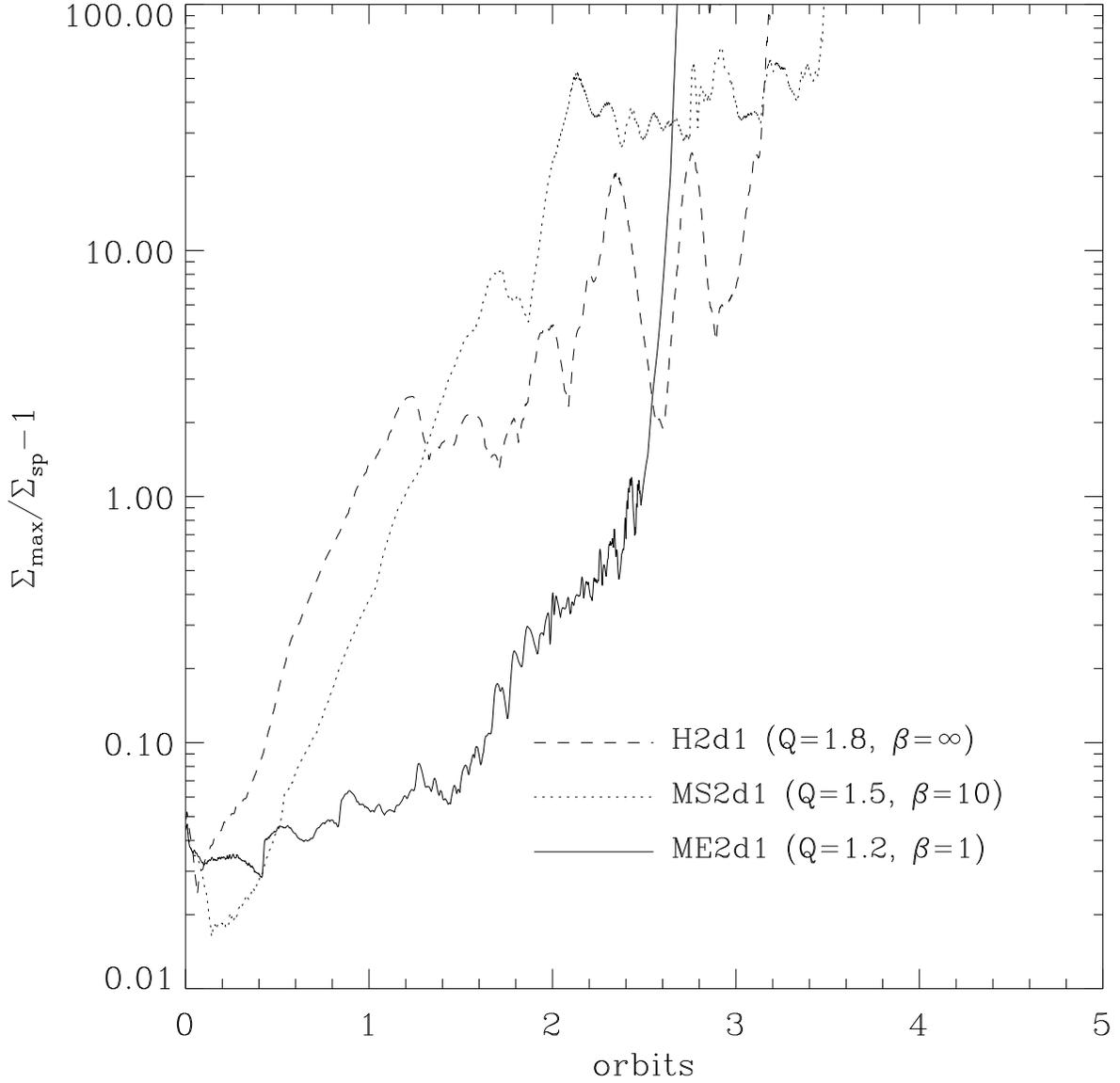}
\caption{Time evolution of maximum surface density for two-dimensional high-resolution 
models. 
\label{evol2D}}
\end{figure}

\begin{figure}
\epsscale{1.}
\plotone{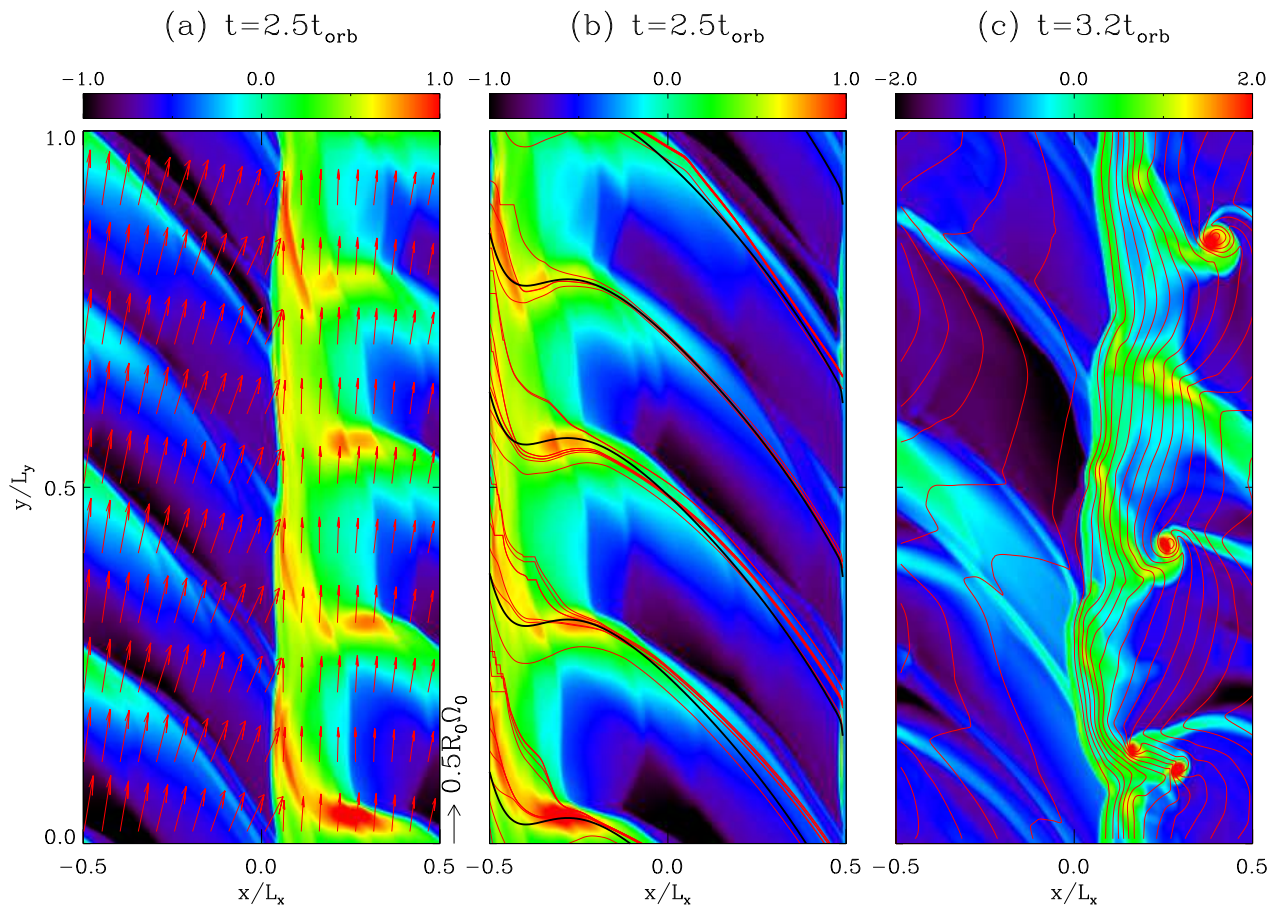}
\caption{Snapshots of model ME2d1 
at $t/\torb=2.5$ and $t/\torb=3.2$. 
For fiducial parameters,
the simulation box has a size of $L_x=L_y/2=3.14$ kpc.
({\it a}) Surface density and
velocity fields seen in the comoving frame with the spiral pattern.
({\it b}) Surface density and a few selected
streamlines ({\it red}) in the frame comoving with the
spurs, together with wavefronts ({\it black})
defined by equation (12) of Paper I
with $K_x(0)=0.45$. The $y$-boundaries are shifted 
so as to make the left wall coincide with the shock location.
({\it c}) Magnetic field lines ({\it red}) are overlaid
on surface density.  Colorbars label $\log \Sigma/\Sigma_0$.
\label{MHD1pat}}
\end{figure}

Figure \ref{evol2D} plots time evolution of the maximum 
surface density of model ME2d1 (together with 
other two-dimensional models with different field strength),
while Figure \ref{MHD1pat} shows snapshots of model ME2d1 
at $t/\torb=2.5$ and 3.2.  
The  perturbations introduced into the flow shear around and relax initially 
($t/\torb\simlt0.4$), and then various modes begin to grow due to 
self-gravity ($t/\torb\simlt2$). 
In the absence of magnetic fields,  growth of perturbations under the 
reversed-shear conditions within the arm can be  
inhibited by Coriolis forces;
these usually reduce
the mass-collecting effect of self-gravity in a rotating system.
However, magnetic tension forces from embedded field lines 
counteract Coriolis forces, so that 
potential vorticity is no longer conserved and perturbations can
grow rapidly (\citealt{lyn66,elm87,kim01}).  
The MJI process works best when shear is weak. The spatially varying sense
of shear inside a spiral arm (reversed, then ``normal'') 
keeps the overall shear rate close to zero,
therefore providing favorable conditions for the development of MJI 
within the arm (Paper I).

Since the perturbations in model ME2d1 initially have low amplitude, 
they remain 
in the linear regime through $t/\torb\sim 1.5$. 
It is not until $t/\torb\approx2$, when perturbations have 
crossed the spiral arm 
twice (since $t_{\rm cross}=\torb/[m(1-\Omega_p/\Omega)]$), that
the most dominant MJI mode emerges and shapes 
the condensed gas flowing into the interarm region into
spur structures.
Figure \ref{MHD1pat}{\it a} shows the surface 
density (in logarithmic scale) and velocity vectors at $t/\torb=2.5$ viewed
from a frame corotating with the spiral arm. 
We identify 4 spurs that protrude, at fairly regular intervals, 
perpendicularly from the main spiral arm, then sweeping back into a
trailing configuration in the interarm region.  
The normalized wavenumber corresponding to
the spur separation $\lambda_y$ in model ME2d1 is 
$K_y=\lambda_{\rm J,sp}/\lambda_y=0.11$.  
These spurs move in the $y$-direction 
with a speed $v_y\sim 0.50 R_0\Omega_0$ relative to the spiral arm, 
implying that in the inertial frame 
they follow very closely the background galaxy rotation 
at the arm center.
Since the lower-resolution model ME2d2 with $128\times 256$ zones also forms 
4 spurs, while model 
ME2d3 with  $128\times 128$ zones results in 3 spurs (see Table \ref{tbl-1}),
we conclude that the number of spurs that form in our two-dimensional
thick-disk models
is independent of numerical resolution as long as 
the $y$-dimension of the simulation domain has 256 zones or more.

Figure \ref{MHD1pat}{\it b} draws selected gas streamlines (in red) seen in
the stationary-spur frame as well as the surface density of
model ME2d1 at $t/\torb=2.5$.  In this figure, coordinates are transformed
such that the left boundary corresponds to the initial location of the 
spiral shock front.  
Because the background flow is shearing and expanding/contracting, the
$x$-wavenumber of Lagrangian perturbations that move away from the shock front
varies as $k_x = - \mathcal{T}k_y$ with
\begin{equation}\label{Wfront}
\mathcal{T} \equiv \frac{1}{\mathcal{R}}\left[
  \frac{\kappa_0^2\Ssp}{2\Omega_0^2\Sigma_0}\tau
  - 2\int_0^\tau \mathcal{R} d\tau - \frac{k_x(0)}{k_y}\right],
\end{equation}
where $\mathcal{R}=\Ssp/\Sigma$ is
the local surface density expansion factor,
$\tau \equiv \Omega_0
\int_{x_{\rm sp}}^x v_{x,T}^{-1} dx$ is a dimensionless elapsed
time that is measured from the shock location (or density peak), $x_{\rm sp}$,
$k_y=2\pi/\lambda_y$ with $\lambda_y$ corresponding to the spur spacing,
and $k_x(0)$ the initial $x$-wavenumber at $\tau = 0$
(Paper I; see also \citealt{bal88}).
Figure \ref{MHD1pat}{\it b} also plots (in black) the theoretical 
wavefronts of spurs given by $dy/dx = -k_x/k_y = \mathcal{T}$
with an initial condition $K_x(0)\equiv  
k_x(0)/k_{\rm J,sp}=0.45$, where 
$k_{\rm J,sp} = 2\pi/\lambda_{\rm J,sp}$.
The fact that the shape of spurs matches quite well with the theoretical
prediction suggests that the former simply reflects the shearing
and expanding properties of the background flow (Paper I).
Note that the gas streamlines rather quickly converge to the spur
wavefront as they move downstream from the spiral shock, indicating
that spurs grow stronger by gathering material mainly along the $y$-direction
(parallel to the spiral arm).

When the density within the spurs has grown sufficiently, 
self-gravity causes them to  
fragment into gravitationally bound condensations.
Figure \ref{MHD1pat}{\it c} shows the density and
 magnetic field lines at $t/\torb=3.2$ of model ME2d1.  
The magnetic fields roughly parallel the arm overall,
although they pinch inward within the spurs and are strongly twisted locally 
in the vicinity of the bound clumps. Bending of field lines is 
most severe in interarm regions where the gas moves faster.
Model ME2d1 forms 4 clumps with mass $M\sim 3.3\times 10^7\Msun$ each;
this is about an order of magnitude larger than the clumps that formed in
razor-thin disk models of Paper I.  Roughly $40\%$ of the total mass, 
therefore, is collected into bound clumps.
These condensations are magnetically supercritical, 
with the mean mass-to-flux ratio
$M/\Phi_B \sim 2.0 G^{-1/2}$, where $\Phi_B$ is the magnetic
flux that passes through each clump and $G$ is the gravitational
constant.\footnote{The critical mass-to-flux limit is $0.16G^{-1/2}$.} 
The numerical box of model ME2d1 initially has a mass-to-flux ratio of 
$M/\Phi_B=3.6 G^{-1/2}$, about twice as supercritical as the clumps that form.
This result is consistent with \citet{vaz05a}, who found that 
self-gravitating substructures formed in turbulent MHD simulations are in 
general less supercritical than the parent system, presumably because 
fragmentation occurring along the flux tubes reduces the cloud mass while 
preserving magnetic flux (see also \citealt{li04}).
Various physical properties of bound clouds that form in two-dimensional 
thick-disk models are quite similar to those in full, three-dimensional models; 
we defer detailed discussion to \S4.2.3, 
where the vertical stratification of disks is explicitly taken into account.

The reduced self-gravity in disks with finite thickness results in a smaller
number of spurs compared to a razor-thin disk with properties otherwise the 
same.  In order to check whether our simulation results are consistent with 
the linear theory prediction, 
we perform a linear stability analysis in a Lagrangian frame comoving with
the background flow through a spiral arm.
Assuming that the perturbed quantities are well described by plane waves with
sinusoidal variations on scales $\ll R$, one can show that 
equations (\ref{con})-(\ref{eos}) and (\ref{kernel}) lead to the 
following set of linearized equations:
\begin{equation}\label{ppcon}
\frac{d\delta\sigma}{d\tau} = K_y(\mathcal{T}\delta u - \delta v),
\end{equation}
\begin{equation}
\frac{1}{\mathcal{R}}\frac{d(\mathcal{R} \delta u)}{d\tau} =
2\delta v - \alpha K_y\mathcal{T}
\left[1 - \frac{1}{\mathcal{R}K(1+KH_0k_{\rm J,sp})}\right]\delta\sigma
- \frac{\alpha \Sigma_{\rm sp}}{\beta_0\Sigma_0} K^2 \delta m,
\end{equation}
\begin{equation}
\frac{d\delta v}{d\tau} =
-\frac{\Sigma_{\rm sp}}{\Sigma_0}\frac{\delta u }{\mathcal{R}}
+ \alpha K_y
\left[1 - \frac{1}{\mathcal{R}K(1+KH_0k_{\rm J,sp})}\right]\delta\sigma,
\end{equation}
\begin{equation}\label{ppind}
\frac{d\delta m}{d\tau} = \frac{\delta u}{\mathcal{R}},
\end{equation}
where $\delta\sigma\equiv \delta\Sigma/\Sigma$,
$\delta u\equiv i \delta v_x k_{\rm J,sp}/\Omega_0$, 
$\delta v\equiv i \delta v_y k_{\rm J,sp}/\Omega_0$,
$\delta m$ is the normalized perturbed vector potential,
$K_y\equiv k_y/k_{\rm J,sp}$, 
$K\equiv |K_y|(1+\mathcal{T}^2)^{1/2}$ is the total wavenumber,
and $\alpha\equiv (c_s k_{\rm J,sp}/\Omega_0)^2$ (see Paper I).

\begin{figure}
\epsscale{1.}
\plotone{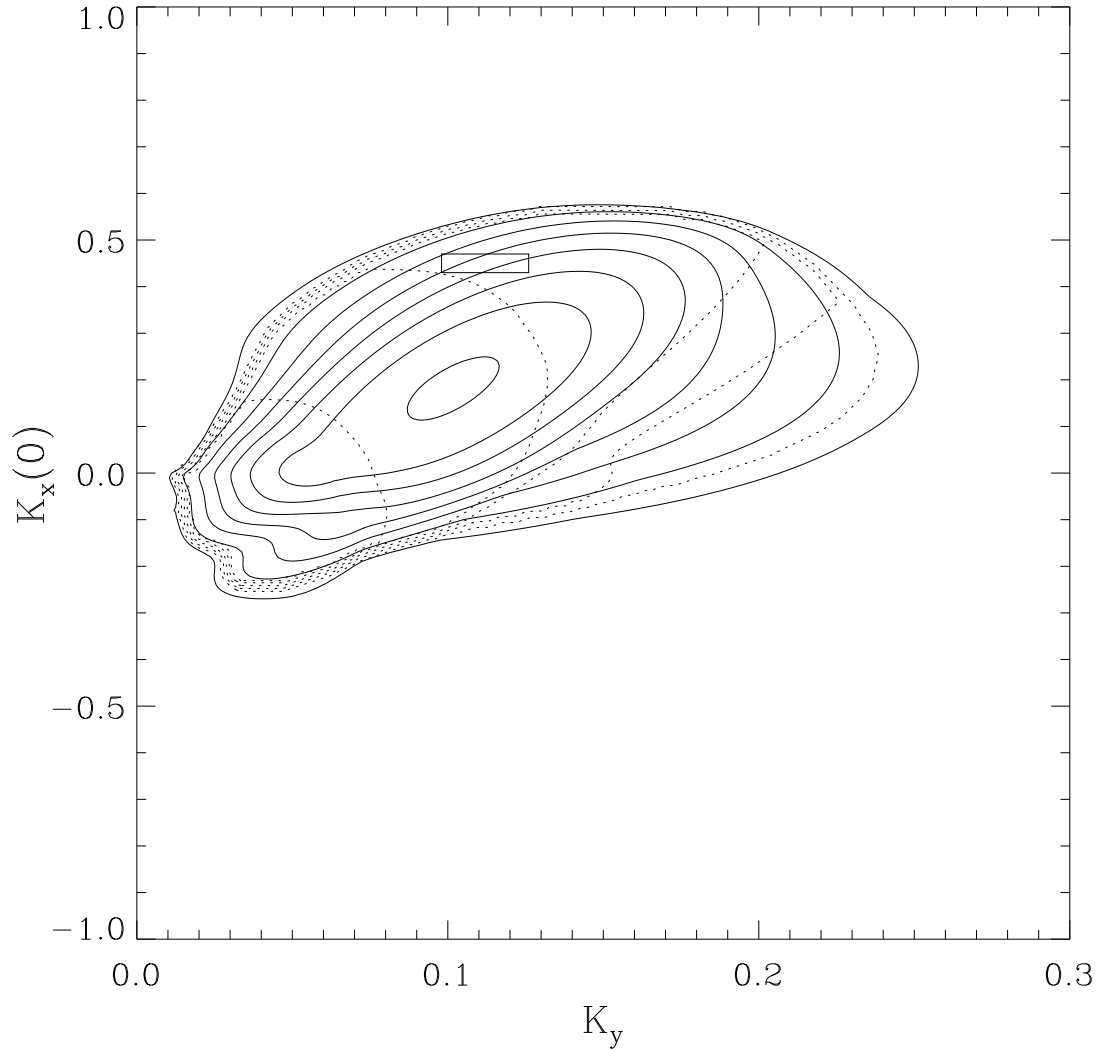}
\vspace{-1cm}
\caption{Prediction of spiral arm MJI development from linear theory, 
with thick-disk gravity, for parameters of 
model ME2d1. The gaseous scale height
$H_0=180$ pc is taken as fixed.  
Solid contours, spaced at $0.4,0.6,0.8,\cdots,1.6,1.8$ from outside to inside, 
show final 
amplification magnitude $\Gamma=\log|\delta\Sigma_{\rm max}/\Sigma|$ 
as a function of $K_y$ and $K_x(0)$, the normalized wavenumbers at the shock 
location parallel and perpendicular to the spiral arm, 
respectively.
Dotted contours plot the corresponding growth times
$\tgrow/\torb=0.2, 0.4, 0.6, 0.8, 1.0$ from right to left.
The most unstable mode evident in the simulation results from 
model ME2d1 is indicated by
the box near $K_y=0.11$ 
and $K_x(0)=0.45$.
\label{linear}}
\end{figure}

We choose $H_0=180$ pc for thick-disk gravity and 
adopt the equilibrium density and velocity profiles (in $x$)
of model ME2d1 as a background state.  By taking 
$\delta\sigma=1$ and $\delta u=\delta v =\delta m=0$ as an initial 
condition and varying $K_y$ and $k_x(0)/k_y$,  we integrate the
perturbed equations in time. 
The resulting amplification 
magnitude $\Gamma\equiv \log |\delta\Sigma_{\rm max}/\Sigma|$ 
and the growth time $\tgrow$ at which maximum amplification occurs
are plotted in Figure \ref{linear} as solid and dotted contours, respectively.
Figure \ref{linear} also marks as a rectangular box
the parameters ($K_x[0]=0.45$ and $K_y=0.11$)
that give the best fit to the shape and spacing of spurs formed 
in the simulation of 
model ME2d1 (the size of the rectangle represents the uncertainty
determining $K_x(0)$ or $K_y$).

Figure \ref{linear} shows that $K_y=0.11$ is in good agreement
with the expectations from linear theory for the wavelength parallel to the 
spiral arm of the dominant mode.  However, the 
spurs that dominate the simulation in 
model ME2d1 
have an $x$-wavenumber at the spiral shock front that is a factor 
2.5 times 
larger than that which would yield 
the largest amplification, based on a linear-theory
analysis.  This is because the initial perturbations in model ME2d1
have such low amplitudes that two traversals of an arm are required before 
the perturbations grow into the nonlinear regime.
Over this time,
$K_x$ of any initial perturbation
varies kinematically due to compression, expansion, 
and shear of the background flow, and overall $K_x$ 
increases due to the large shear in the interarm 
region.
Perturbations that are amplified (via MJI) 
during a first passage through a spiral 
arm become preferentially trailing with high $K_x(0)$
before  they 
enter the next spiral shock (see Figure \ref{MHD1pat}{\it a}).
Because of the strong  compression immediately behind the shock, 
$K_x(0)$ of these trailing wavelets is
further increased before they enter the next stage of amplification 
by MJI.  
Consequently, the perturbations that 
eventually grow to emerge as spurs in model ME2d1 
have quite a large value of $K_x(0)$.

\begin{figure}
\epsscale{1.}
\plotone{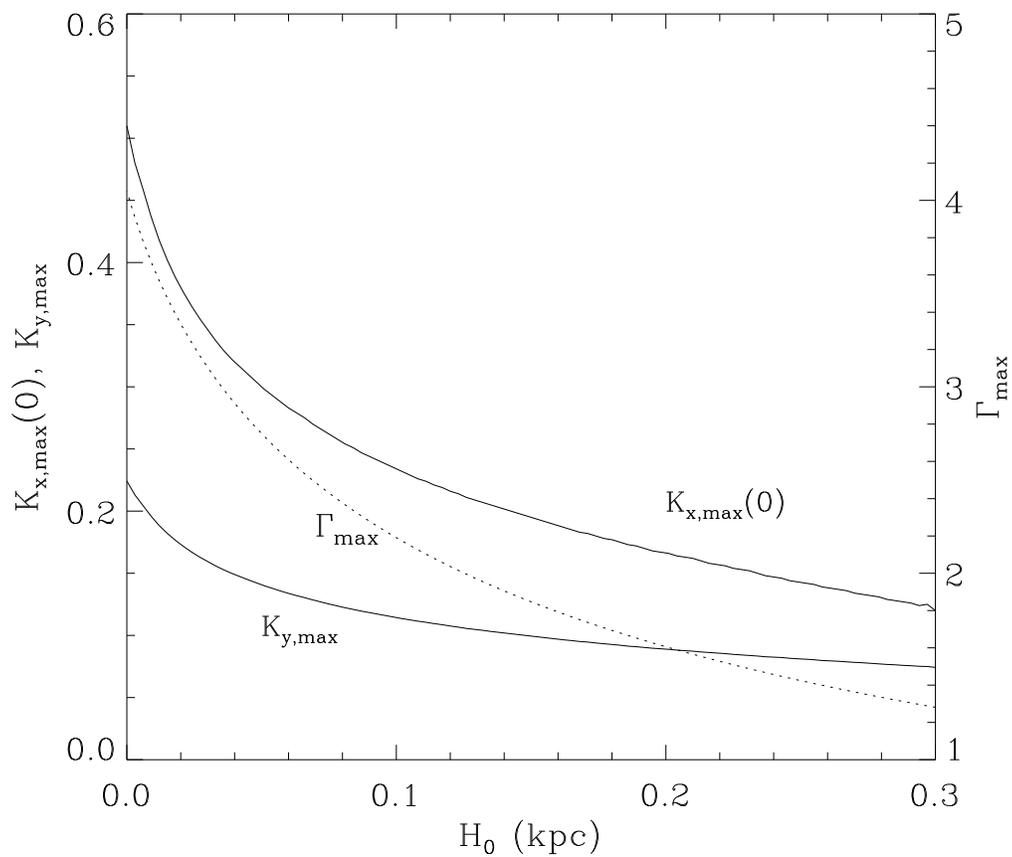}
\caption{Dependence on the gaseous scale height $H_0$ of
(solid lines; left $y$-axis) the wavenumbers $K_{x,\rm max}(0)$ and 
$K_{y, \rm max}$ and (dotted line; right $y$-axis) the amplification 
magnitude $\Gamma_{\rm max}$ of the  dominant MJI mode in 
spiral shocks predicted from linear theory.  The 
parameters for 
model ME2d1 are used.
\label{amp_H}}
\end{figure}

To quantify how finite disk thickness is expected to affect
the development of MJI  inside
spiral arms, we vary $H_0$ and search for $K_{x,\rm max}(0)$ and 
$K_{y,\rm max}$ that give -- based on linear theory -- 
the maximum amplification magnitude 
$\Gamma_{\rm max}$ for a given value of $H_0$.
Figure \ref{amp_H} plots the resulting $K_{x, \rm max}(0)$ and 
$K_{y, \rm max}$ as solid lines as well as $\Gamma_{\rm max}$ as a dotted line.
The equilibrium surface-density profile of model ME2d1 
(but with varying $H_0$)
is again used.  
The linear theory suggests that the MJI inside spiral 
arms in a disk with $H_0=180$ pc requires $K_{y, \rm max}$
to be 2.6 times smaller than in a disk with $H_0=0$.  
The corresponding amplification magnitude $\Gamma_{\rm max}$ is
2.3 times smaller for thick disks compared to thin disks.  For comparison,
we note that in the razor-thin models of Paper I,
spurs identified in simulations with $\beta_0=1$ typically have 
$K_y\sim 0.45-0.54$ 
(or $\lambda_y \sim 1.8-2.2 \lambda_{\rm J,sp}$), 
with larger $K_y$ (or smaller $\lambda_y$) 
corresponding to smaller-$Q_0$ models. 
Since the amplification factor is enormous when $H_0=0$,
the modes that come to dominate in razor-thin disks tend to be 
selected by compromise between maximum amplification 
and earliest growth (Paper I), and thus have $K_y$ larger than that 
corresponding to $K_{y, \rm max}$.

In the thick-disk models with their lower amplification, on the other hand, 
early growth gives little advantage over other modes.
Model ME2d1 therefore chooses the mode corresponding to the largest total 
amplification, 
whose wavelength turns out to be $\lambda_y\sim 9 \lambda_{\rm J,sp}$ which
is $\sim 4-5$ times larger (in terms of 
$\lambda_y/\lambda_{\rm J,sp}$) than that found for razor-thin disk models 
in Paper I.
Models with thick-disk gravity have peak surface densities
twice as large as their razor-thin counterparts with the same $F$ and
$Q_0$, which reduces the Jeans
length by about a factor of two.  
Thus, for cloud mass scaling as 
$\lambda^2 \Sigma_{\rm sp} \propto (\lambda_y/\lambda_{\rm J,sp})^2
\Sigma_{\rm sp}^{-1}$, 
one can expect that bound clouds resulting from 
spur fragmentation in thick-disk gravity models should be $\sim 8-10$ times 
more massive than in razor-thin disks with the same $F$ and $Q_0$, or 
$\sim(3-4)\times 10^7\Msun$ in dimensional units.
This is entirely consistent with the results of our numerical simulations. 

\subsection{The Wiggle Instability}

We now turn to model MS2d1 (and its lower-resolution counterparts
MS2d2-3),
 with $Q_0=1.5$, $F=5\%$, and $\beta_0=10$ 
(i.e., $\vA=\cs/\sqrt{10}$), and unmagnetized model H2d1 (with
counterparts H2d2-3) having
$Q_0=1.8$, $F=5\%$, and $\beta_0=\infty$.  
As Figure \ref{evol2D} shows, in models MS2d1 and H2d1 perturbations
immediately begin to 
grow almost exponentially over time, while strong growth in 
model ME2d1 
does not begin until after 
two successive passages through spiral arms.
The growth rates of the maximum
surface density in models MS2d1 and H2d1 are almost identical,
$\sim0.70\Omega_0$, suggesting that moderate magnetic fields are not
important at least in the linear stages of growth in these simulations.

\begin{figure}
\epsscale{1.}
\plotone{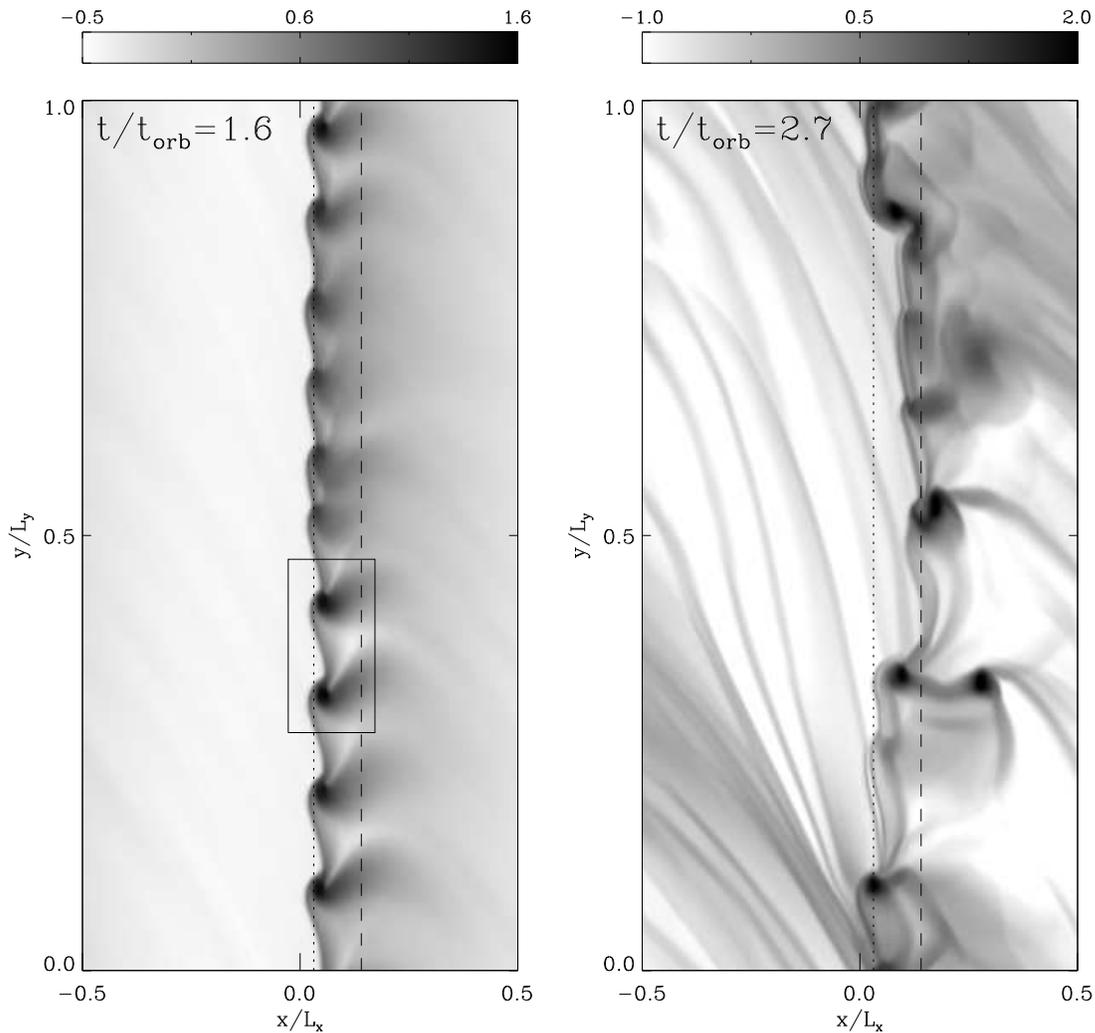}
\caption{Surface density maps ($\log(\Sigma/\Sigma_0)$ in gray scale) 
of model MS2d1 
at ({\it left}) $t/\torb=1.6$ and ({\it right}) $t/\torb=2.7$. 
For fiducial parameters,
the simulation box has a size of $L_x=L_y/2=3.14$ kpc.
In both panels, the initial shock location is indicated by the dotted line, 
while the dashed line marks the position where the sense of shear 
in the initial background flow changes from reversed to normal.
The rectangular section in the left panel is enlarged in Figure \ref{wiggle}
to show its temporal changes. 
\label{MHD10_2d}}
\end{figure}

The two models MS2d1 and H2d1 have similar evolution, but this is 
quite distinct from the behavior of equipartition-magnetization
 model ME2d1.
Unlike model ME2d1, in which the spiral shock remains relatively
straight during the initial phase of MJI growth, 
models MS2d1 and H2d1 show  small-scale instability that
wiggles the spiral shock front and forms structures with high vorticity. 
The left panel of Figure \ref{MHD10_2d} plots the surface density map 
in logarithmic gray-scale of model MS2d1 at $t/\torb=1.6$. 
The dotted line marks the location of the spiral shock front, 
while the dashed line represents
the position where shear in the background flow vanishes ($dv_{T,y}/dx=0$).
Evidently, the vortical clumps that emerge are 
closely connected to the shock discontinuity.
Column (9) in Table \ref{tbl-1} shows that 
the hydrodynamic models H2d1-3 form more clumps than 
the weakly magnetized models MS2d1-3,
and that the number of vortical clumps produced depends rather sensitively 
on numerical resolution.

At $t/\torb\approx 1.7$, the growth of vortical structures in model MS2d1 
saturates due to magnetic tension from bent field lines as well as 
nonlinear effects.
(The flow properties
inherent in the background spiral shock may also prevent the
clumps from growing further.) 
At this time, the clumps are not sufficiently self-gravitating to 
experience gravitational collapse due to their own weight.  The clumps
wander slightly  inside the postshock
regions and collide with each other to form bigger clumps.  
As Figure \ref{MHD10_2d}$b$ shows, mergers of 
10 low-density clumps at $t/\torb=1.6$ result in 5 high-density, 
gravitationally bound clumps at 
$t/\torb=2.7$.  These bound clumps have an average mass
$\sim1\times 10^7\Msun$, a factor of 3 smaller than that originating 
from MJI in model ME2d1, and are supercritical with an average mass-to-flux 
ratio $M/\Phi_B\sim 3G^{-1/2}$.  
The interarm spur structures prominent 
in model ME2d1 (see Fig.\ \ref{MHD1pat}) are not clearly visible in models 
MS2d1 and H2d1. 

\begin{figure}
\epsscale{1.}
\plotone{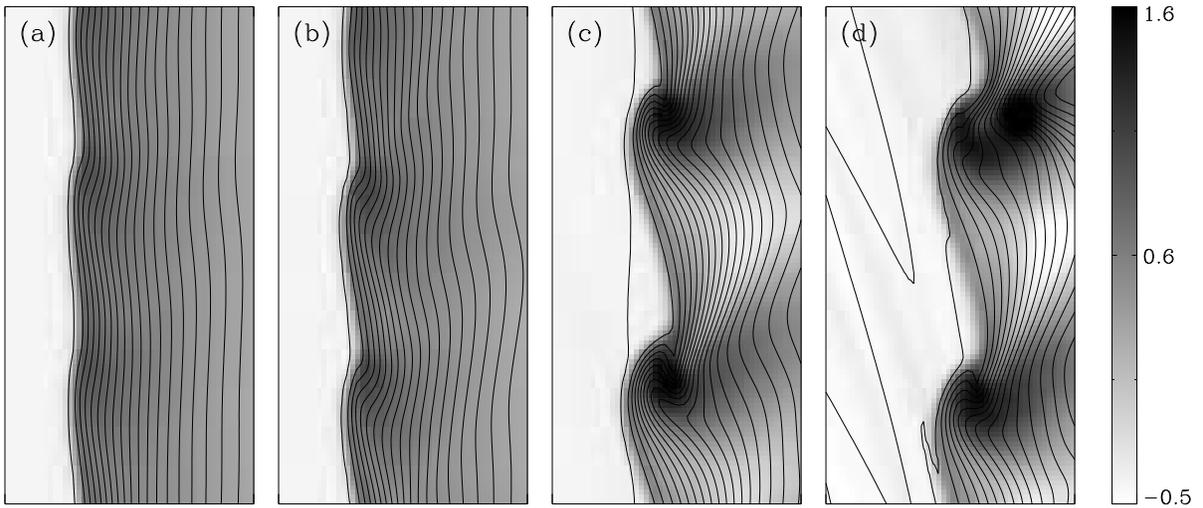}
\caption{Temporal changes of the rectangular section shown in the left panel 
of Figure \ref{MHD10_2d} at
$(a)$ $t/\torb=1.1$,
$(b)$ $t/\torb=1.3$,
$(c)$ $t/\torb=1.6$, and
$(d)$ $t/\torb=1.9$.  In each panel, magnetic field lines are overlaid
on surface density ($\log(\Sigma/\Sigma_0)$ in gray scale).
\label{wiggle}}
\end{figure}

What causes the spiral shocks to wiggle and form vortical structures
in models MS2d1 and H2d1?
Figure \ref{wiggle} shows the temporal evolution of gas surface density and 
magnetic field lines in the rectangular section shown in the left frame of
Figure \ref{MHD10_2d}.
Note that the vortex generation and evolution of the magnetic field 
topology in model MS2d1 are similar to those that result from Kelvin-Helmholtz 
instabilities in unmagnetized (e.g., \citealt{cor84}) or
magnetized (e.g., \citealt{mal96,fra96}) shear layers.  
This suggests that the vortical clumps in 
model MS2d1 may arise from  Kelvin-Helmholtz instability in the sheared
flow -- i.e. azimuthal streaming --  associated with spiral shocks. 
\citet{wad04} performed non-self-gravitating 
global simulations of razor-thin, unmagnetized
disks and found that when spiral perturbations are strong and have 
large pitch angles, the shock front that develops 
wiggles and forms discrete clumps 
bearing remarkable resemblance to those found in our models.
Wada and Koda termed this process a  ``wiggle instability''.
Based on  the low Richardson numbers in the postshock regions,
they argued that the formation of these clumps 
could be due to  Kelvin-Helmholtz instability.
As Wada and Koda noted, however, the Richardson number criterion is 
only a {\it necessary} condition for {\it stability} 
\citep{cha61}, and should not be regarded as an 
instability criterion.  The critical value of Ri may, in addition, be affected
by rotation.  Thus, while it is likely that the wiggle instability
is indeed a manifestation of 
 Kelvin-Helmholtz instability in the shear flows produced by
spiral shocks, it is not yet certain.

Since models considered in \citet{wad04} are non-self-gravitating and
unmagnetized, the wiggle instability, whether it is related to
Kelvin-Helmholtz instability or not, relies neither upon magnetic fields
nor self-gravity.  Paper I found that gas in razor-thin disks 
with $F\leq3\%$ is subject to either the MJI or the swing amplifier, but
is stable to the wiggle instability; i.e., non-self-gravitating models 
at these values of $F$ are expected to be stable. 
We have run a number of two-dimensional simulations
of non-self-gravitating,  unmagnetized disks with varying $F$
(not listed in Table \ref{tbl-1}), and found that the wiggle instability 
becomes manifest when $F\geq4\%$.
This suggests that spiral shocks must be fairly strong in order to trigger
the wiggle instability\footnote{\citet{wad04} in fact
set up extremely strong spiral shocks whose amplitudes amount to
$F\approx2\epsilon\cot i\sim110\%$ for the parameters 
$\epsilon=0.1$ and $i=10^{\rm o}$ in their model A; the
arm/interarm contrasts reached two orders of magnitude.}.
Since non-self-gravitating gas behind a spiral shock is {\it locally} 
Kelvin-Helmholtz 
stable due to the expanding radial velocity inside the spiral 
arm \citep{dwa96},  the wiggle instability is most likely a 
consequence of interaction between preshock gas and compressed 
postshock flows 
that stream at large relative velocity parallel to the shock front.
It is also possible that spiral shocks are so strong that 
the postshock velocity perpendicular to the arm is vanishingly small,
limiting the stabilization to Kelvin-Helmholtz modes.  
In  any event, we do not believe the wiggle instability is likely to be
important in spiral arms of 
real galaxies, because as we will show in the next section it does not occur
in three-dimensional models, in which the fluid variables are not constant 
with height.

\section{Three-dimensional Models} 

To investigate the effects of disk stratification and other vertical
variations on 
the formation of spiral-arm substructures,
we have performed three sets of three-dimensional numerical simulations.
The chosen parameters for the three-dimensional models listed in Table \ref{tbl-2} are
identical to those of the two-dimensional models presented in the previous section.
We first set up an axisymmetric disk with equilibrium
density distribution $\rho_0(z)$ 
consistent with a chosen set of 
$Q_0$, $\beta_0$, and $s_0$.  We then apply initial perturbations
to the background density $\rho_0(z)$, using a spatially 
uncorrelated, Gaussian random field that has flat power 
for $1\leq kL_x/2\pi\leq64$ and zero power for $64< kL_x/2\pi$.  
For the amplitude of the perturbations, we fix the standard deviation of 
perturbed density in real space to be 3\% of the initial midplane density. 
Next, we slowly turn on a spiral potential perturbation 
(the second term in eq.\ [\ref{ext_pot}]) such that it acquires the
full strength $F$ at $t/\torb=1.5$.

\subsection{Vertical Structure of Spiral Shocks}

\begin{figure}
\epsscale{1.}
\plotone{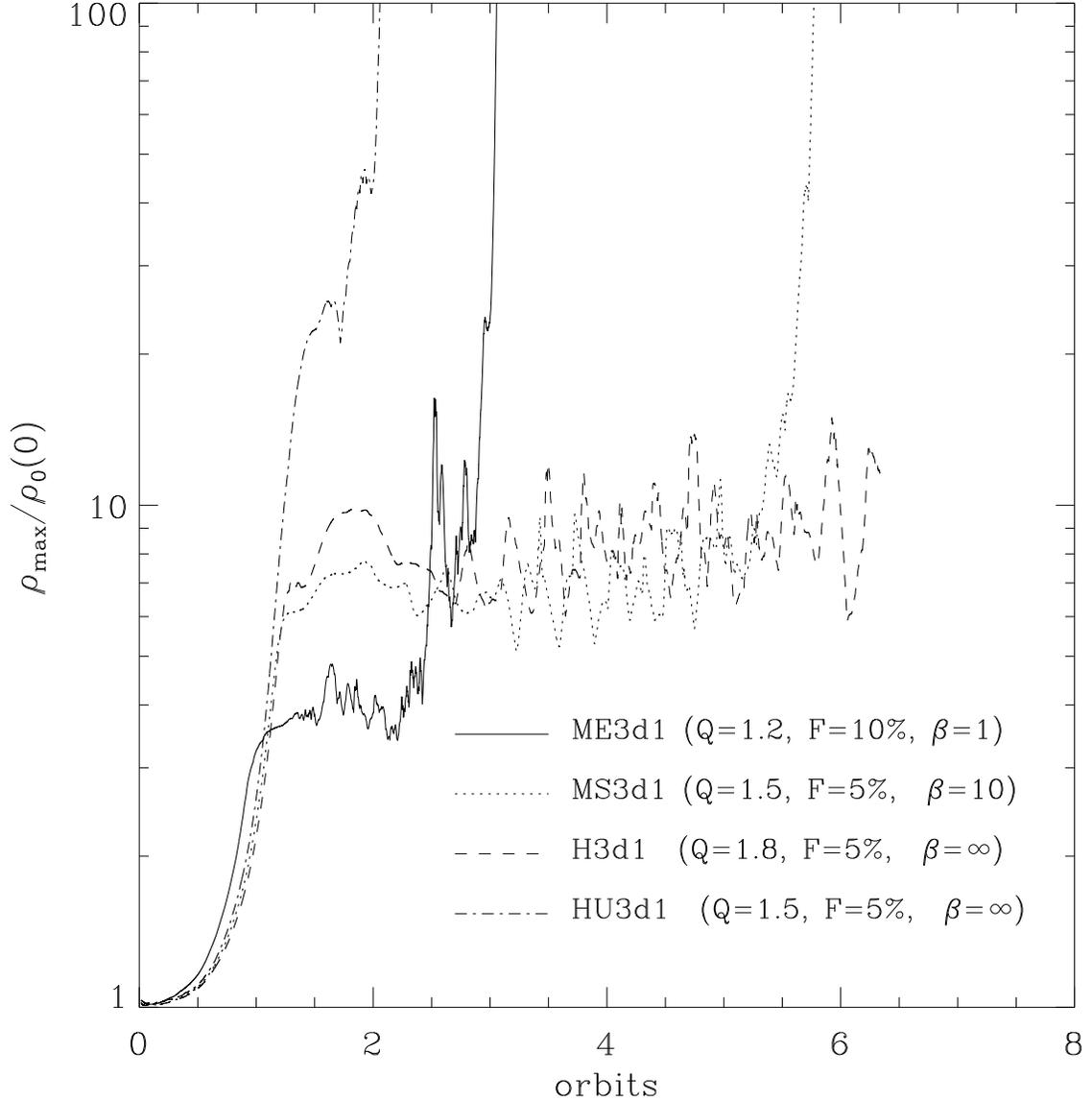}
\caption{Maximum gas density  in units of the initial midplane density 
$\rho_0(0)$ for high-resolution three-dimensional models.  
The imposed spiral potential reaches full
strength at $t/\torb=1.5$.  
While the density in unmagnetized model H3d1
fluctuates around a mean value and does not become sufficiently
nonlinear for gravitational instability, 
magnetized models ME3d1 and MS3d1, which are subject to MJI,
evolve into a highly nonlinear state, forming 
spurs and gravitationally bound condensations.
Unlike models ME3d1, MS3d1, and H3d1 which permit stable shock equilibria,
model HU3d1 produces a spiral shock that grows exponentially via
quasi-axisymmetric Jeans instability and subsequently experiences
non-axisymmetric fragmentation to form
bound condensations at the shock front.
\label{evol3D}}
\end{figure}

Figure \ref{evol3D} shows evolution of the maximum density in 
high-resolution three-dimensional models ME3d1, MS3d1, and H3d1. 
As the strength of the spiral potential increases, a density wave grows
rapidly in the gas, until the initially-sinusoidal profile
steepens into a shock.  During this phase, the growth
of small-scale nonaxisymmetric perturbations is weak 
 compared to the large-scale 
nonlinear response of the gas disk  to the imposed 
spiral potential perturbation.
At  $t/\torb\sim1.6-1.8$, slightly after the spiral potential attains
its full strength, the spiral shock reaches its maximum strength,
leaving density and velocity fields that fluctuate around 
the respective average configuration.
At this stage,
the spiral shock still remains almost axisymmetric.

Paper I and \S3 of the present paper showed that when vertical degrees of
freedom are ignored, quasi-axisymmetric, steady-state profiles of gaseous 
spiral shocks
can be readily obtained through time-dependent calculations 
in an external spiral potential (see also \citealt{woo75}).
This implies that one-dimensional spiral shock profiles in the
galactic midplane represent {\it stable} shock equilibria
(provided that gaseous self-gravity and external spiral forcing 
are not too strong; see Paper I).  When we allow for 
vertical degrees of freedom in the fluid
variables, however, we find that the profiles that develop are not in general
quasi-steady.  
In the initial stages of all the 
three-dimensional models studied 
in this work, quasi-axisymmetric, two-dimensional spiral shocks develop in the
$x$-$z$ plane (hereafter XZ spiral shocks) that are in general non-stationary, 
swaying loosely back and forth in the direction perpendicular to the arm. 

Based on our simulation results, the
 ``flapping'' motions of  XZ spiral shocks are strongest at high altitudes,
and stronger in model ME3d1 than in models MS3d1 and H3d1. 
We measure the maximum  shock front excursion from
its mean location as about 0.06$L_x$, $0.02L_x$, and $0.03L_x$ at $z/H_0=1$
in models ME3d1, MS3d1, and H3d1, respectively. 
The non-steady nature of XZ spiral shocks that we find 
is in fact commonly seen in other
numerical models (e.g., \citealt{mar98,gom04}). 
Although \citet{sou81} constructed stationary shock profiles in
a vertically extended disk, their models neglected 
the effects of the Coriolis force, self-gravity, and magnetic fields.
The flapping motions of XZ spiral shocks may just be overshooting of
gas caused by instantaneous force imbalance across the shock, or
may represent a dynamical instability. 
Detailed discussion on the physical origin of the
spiral shock flapping, and the characteristics of the associated velocity
and density fields, will be presented in a subsequent paper.

\begin{figure}
\epsscale{0.9}
\plotone{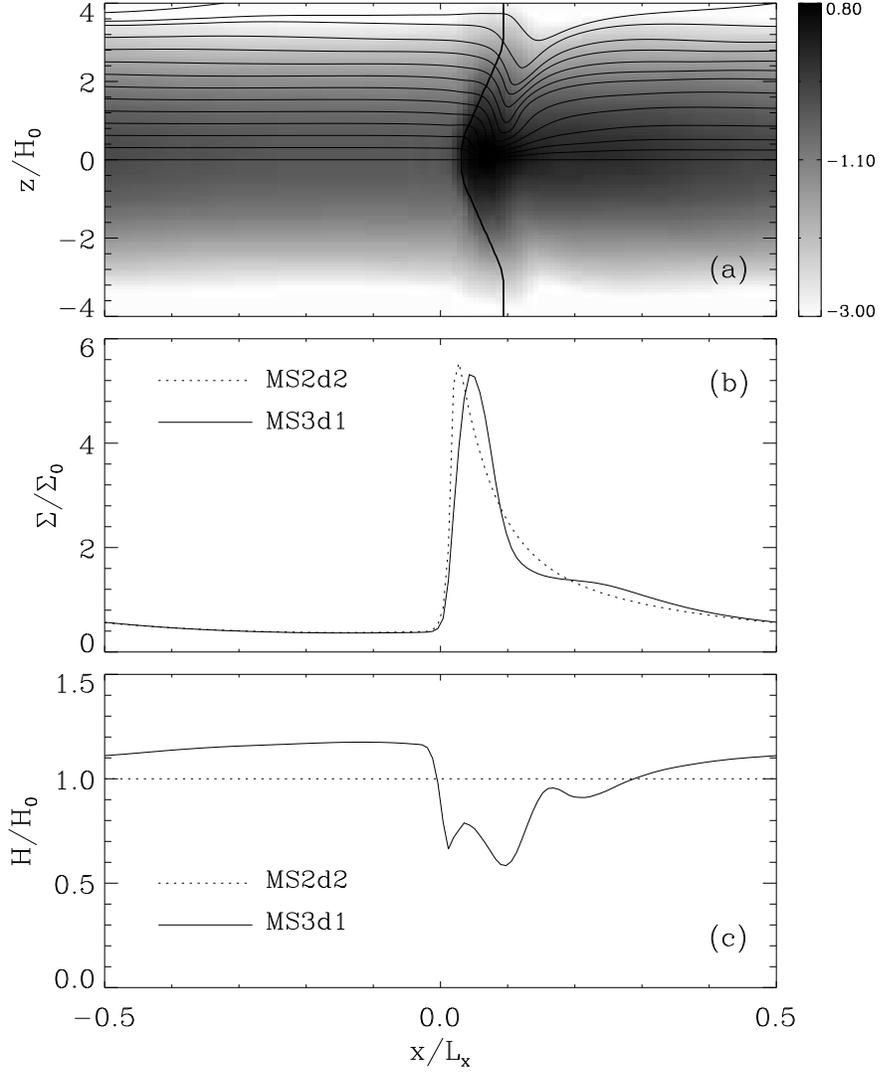}
\caption{Mean spiral shock solutions 
 in the radial-vertical (XZ) plane of model MS3d1,
based on spatial average along azimuth ($y$)
and time average over $t/\torb=1.7-2.7$.
({\it a}):  Gas streamlines (thin solid lines) are overlaid on 
the gas density ($\log(\rho/\rho_0)$ in gray scale) .  The heavy vertical
 line  marks the  averaged shock front.
({\it b}): Average 
surface density profile (solid line) based on vertical integration of
density in  
(a) is compared to its two-dimensional thick-disk counterpart from
model MS2d2 (dotted line).
({\it c}): The profile of mean gaseous 
scale height 
for models MS3d1 (solid line) and MS2d2 (dotted line).  
The gaseous scale height is generally smaller inside spiral arms
than in interarm regions.
\label{equil_xz}}
\end{figure}

Since spiral shocks in three dimensions exhibit temporal
fluctuations, it is useful to take space-and-time averages in order to
visualize the shock structures in the $x$-$z$ plane.  Figure
\ref{equil_xz} illustrates this using model MS3d1, with the density
and velocity fields averaged along the $y$-direction and also over an
orbital period from $t/\torb=1.7$ to 2.7.  Figure \ref{equil_xz}$a$
shows gas density in logarithmic scale and selected streamlines.  The
streamlines run almost parallel to the galactic midplane in the
interarm region, but sharply plunge toward the midplane in the arm.
The location of the averaged shock front is also indicated by a heavy line. 
The low vertical velocities of the
gas in the interarm regions implies that 
vertical force balance is fairly well
maintained there.  As the material enters the spiral arm region, it
is shocked and compressed.  The increase of gas density due to the
shock compression is largest at the midplane, which in turn produces
strong vertical gravity and pulls high-altitude gas toward the midplane.  The
vertically-plunging gas further increases the midplane density after the
shock, setting up
repulsive pressure gradients that cause it to rebound back
to high-altitude regions.  The furrow in the  streamlines behind the
shock front reflects this vertical dive and bounce.

Figure \ref{equil_xz}{\it b,c} plots as solid lines the surface 
density profile $\Sigma(x)=\int_{-\infty}^\infty\rho(x,z)dz$ and the 
corresponding gas scale height $H(x)=\Sigma(x)/(2\rho[x,0])$ obtained 
from the averaged 
XZ spiral shock shown in Figure \ref{equil_xz}{\it a}.
Also plotted as dotted lines are the $\Sigma(x)$ and $H(x)$ values 
of the thick-disk two-dimensional 
model MS2d2 that has 
exactly the same parameters and  numerical resolution as model MS3d1, 
but fixed $H_0=170$ pc. 
Although the surface density in model MS2d2 rises slightly more steeply at the
shock than in model MS3d1, they are overall in excellent 
 agreement.
This proves that the gravitational kernel given by equation (\ref{kernel}) 
accurately handles the dilution of self-gravity in an extended disk.

As Figure \ref{equil_xz}{\it c} shows, the gas within spiral arms
is more compressed toward the midplane than in interarm regions. 
The tendency for  gas to compress vertically in the arms 
increases as  the magnetic field strength  
 decreases. Column (8) of Table \ref{tbl-2} indicates that the 
ratio of arm to interarm scale heights is about
0.5 for models with $\beta_0=\infty$ or 10, and 0.8 for $\beta_0=1$ models.
This is because models with 
weaker magnetic fields have higher arm density enhancement,
and thus relatively stronger self-gravity that vertically
compresses spiral arm gas.  We discuss the relationship of this
finding to observations in \S 5.2.

Finally, we remark that our models do not exhibit the ``hydraulic
jump'' behavior that \citet{mar98} suggested may develop, under
certain circumstances, at spiral shock fronts.  While hydraulic jumps
(which yield an increase in the gas scale height where it is
compressed in arms) can occur when the equation of state is fairly
stiff, \citet{mar98} note that when $\gamma=1$ there are only shocks.
Magnetic fields parallel to the arm may provide extra stiffness,
but self-gravity tends to draw gas toward the midplane in the
post-shock region; in our (isothermal) models this results in a
decrease of $H$ within arms.  It is not presently known whether the
tendency of self-gravity to reduce $H$ would overcome the tendency of 
a sufficiently stiff equation of state to increase $H$ in arms, since 
the studies of \citet{mar98} did not include self-gravity.

\subsection{Nonaxisymmetric Evolution of Three-Dimensional Disks}

\subsubsection{Absence of the Wiggle Instability in Three-Dimensional Disks}

Based on our two-dimensional thick-disk models, we found that perturbations
in spiral shocks can grow, either as a result of the MJI or the wiggle
instability. One may naively expect that three-dimensional models would also be
prone  either to MJI or to wiggle instability, 
not to mention to other potential three-dimensional instabilities.
Surprisingly, however, we find that the wiggle instability seen in some of our
two-dimensional models is absent in three-dimensional models, while MJI still 
operates in three-dimensional models. 
As we will show below, our three-dimensional model MS3d1 with $\beta_0=10$ is 
subject to MJI, whereas the corresponding two-dimensional 
model MS2d2 (with identical in-plane resolution to model MS3d1)
was found to be unstable to the wiggle instability.  In addition, the
unmagnetized two-dimensional models H2d1-H2d3 developed vortical clumps
via the wiggle instability that later merged and collapsed, but
as Figure \ref{evol3D} shows, the corresponding three-dimensional model H3d1 
does not develop strongly overdense clumps over the entire run.
The peak  density in model H3d1
fluctuates with period $\sim0.29\torb$, comparable to 
the period associated with epicyclic motion, 
$\sim (2\Sigma_{\rm sp}/\Sigma_0)^{-1/2}\torb\approx0.31\torb$, within the arm
peak.

\begin{figure}
\epsscale{1.0}
\plotone{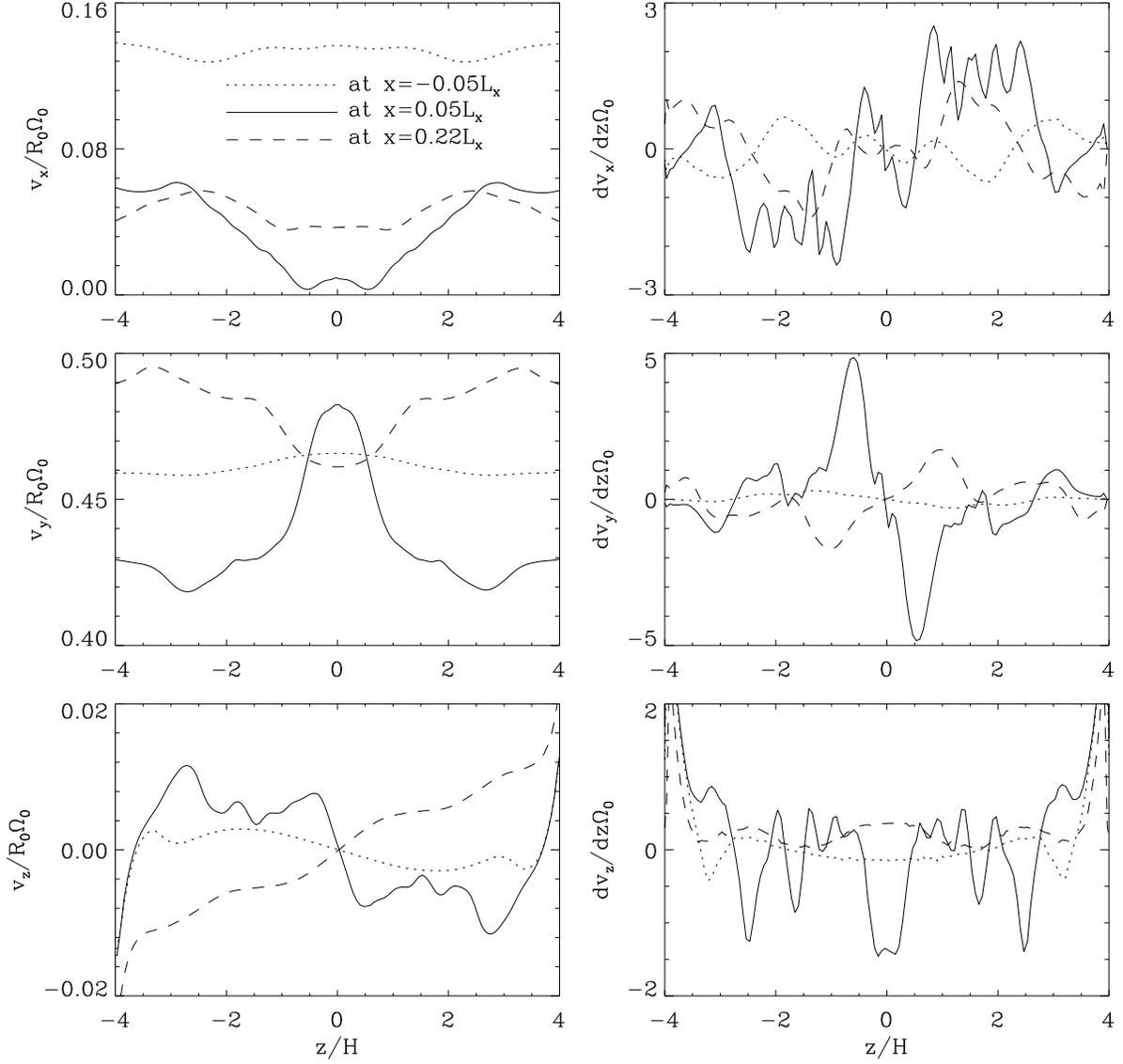}
\caption{({\it left}) Vertical distributions and ({\it right}) gradients 
of spiral shock velocities.  Measurements are
 slightly upstream from the midplane
shock 
at $x/L_x=-0.05$ (dotted), slightly downstream from the shock at
$x/L_x=0.05$ (solid), and in the interarm region at $x/L_x=0.22$ (dashed).
Profiles are based on the averaged XZ shock configuration of model MS3d1. 
Shear in the horizontal velocities is very strong for 
$0.2\simlt|z/H|\simlt2$.
\label{vshear}}
\end{figure}

The wiggle instability seen in two-dimensional disks thus appears to be 
stabilized in three-dimensional disks, possibly due to a combination of 
strong vertical shear
and non-steady flow.  Figure \ref{vshear} shows the vertical
distribution and gradients of the azimuthal- and time-averaged spiral
shock velocities at $x/L_x=-0.05$ (upstream), 0.05 (at the midplane
density peak), and 0.22 (far downstream), from model MS3d1.  The
velocity shear along the vertical direction amounts to $q_{xz}\sim 1$
and $q_{yz}\sim5$ at $|z/H|\sim0.5$ and $q_{xz}\sim 2$ at
$|z/H|\sim1-2$, where $q_{xz}\equiv \Omega_0^{-1}|dv_{x}/dz|$ and 
where $q_{yz}\equiv \Omega_0^{-1}|dv_{y}/dz|$ . 
These shear rates can be compared
with $q_0=1$, the radial shear rate in the background azimuthal flow, and 
with the maximum radial shear in the quasi-azimuthal flow within the 
arm, $q=|2-(2-q_0)\Sigma_{\rm sp}/\Sigma_0|\approx 4.3$.
Perturbations in the spiral shock of 
model MS3d1 might try to develop nonaxisymmetric vortical flows
similarly to those produced by the wiggle instability 
in two-dimensional thick-disk models.  
However, these vortical structures 
could not remain coherent in $z$ because of the strong vertical shear in 
horizontal velocities; vertical motions would then mix dissimilar vorticity, so
that the vortices would be lose their integrity.  In addition, non-steady
flapping of the shock front would also help prevent coherent structures
from forming.  
Overall, the various three-dimensional effects combine to suppress nonlinear
development of the wiggle instability.

Since our three-dimensional models explore cases with $F=5-10\%$, our results 
are not directly comparable to cases in which spiral shocks are as strong as 
in \citet{wad04}'s models.  Strong spiral shocks in two dimensions are more 
favorable for promoting wiggle instability, but in three dimensions they would 
also result in stronger shock flapping motions and stronger vertical shear of 
the in-plane velocities, which tend to suppress the wiggle instability. 
Thus, it would require direct three-dimensional simulations with very large $F$
to determine whether increasing shock strength overall stabilizes or 
destabilizes spiral shocks to ``wiggle'' modes. 
However, we consider this question of limited practical importance, since real 
astronomical systems would in general become self-gravitating if very strong 
shocks were present.  In this sense, the non-self-gravitating treatment of 
\citet{wad04}, with $F\sim110\%$, appears unrealistic given that the
spiral 
shocks in their models have density jumps of a factor 100.  
Contrasts in real galaxies are far smaller; e.g.  the notoriously strong 
spiral structure in M51 has arm/interarm contrasts of 5-6 
(\citealt{gar93}; see also \citealt{rix95,pat01}).  
This is because equilibrium shock profiles are only possible at moderate 
values of $F$ (Paper I).  For instance, when $Q_0=1.8$ and $\beta=\infty$, 
quasi-axisymmetric shock equilibria with thick-disk gravity exist only if 
$F \simlt 20\%$, beyond which shock profiles are highly transient or suffer 
from immediate quasi-axisymmetric gravitational instability 
(see section 4.2.4).

\subsubsection{Swing Amplification}

When shear is strong (with $q\simgt 0.3$) 
and magnetic fields are absent, nonaxisymmetric modes 
are not subject to MJI but can grow via swing amplification 
\citep{gol65b,too81,kim01}. 
The growth of wave amplitudes via swing amplification is moderate unless 
the $Q_0$ value of the background medium is sufficiently small.
\citet{kim01} showed that swing amplification in a razor-thin, featureless 
disk with $Q_0\sgreat 1.3$ yields low  density fluctuations
and cannot form gravitationally bound clumps.
Paper I further found that the swing amplifier 
produces substructure growth 
in spiral arms only if the background spiral 
perturbation is relatively weak (with $F \sles1\%$ in a razor-thin disk).
In this case, the density enhancement in the shock is quite low 
and all the gas has normal shear,
as opposed to the locally-reversed shear which occurs wherever 
$\Sigma/\Sigma_0>2$.

With $F=5\%$, model H3d1 in the present work has strong shear reversal 
corresponding to $q_{\rm min}=-4$  inside a spiral arm.
In addition, a relatively large value of $Q_0=1.8$ as well 
vertically-diluted 
self-gravity make swing amplification in model H3d1 quite ineffective.
The resulting nonaxisymmetric structures in the surface density of 
model H3d1 at the end of the run ($t/\torb=6.4$) 
have the maximum power in the $\lambda_y=L_y/3$ mode, which is
only $\sim0.26\%$ relative to the axisymmetric mode;
the associated maximum surface density is 1.46 times 
the mean surface density.  We thus conclude that the swing
amplification mechanism
is 
unlikely to prompt the formation of substructure within spiral arms
in real disk galaxies where spiral arms are not so weak.  Of course, as the 
models of Paper I showed, even without magnetic fields, if $Q_0$ is small
enough, and/or if $F$ is large enough, 
spiral arms can become unstable to fragmentation.  This fragmentation
in general {\it first} develops via quasi-axisymmetric instability, so the 
process is physically distinct from direct swing amplification.  We will 
discuss this alternative fragmentation process in \S \ref{AxiU}.

\subsubsection{Magneto-Jeans Instability in Three-Dimensional Disks}

In the preceding subsections, we have shown that the wiggle
instability disappears in three-dimensional extended disk models, and that 
swing amplification is only moderate in unmagnetized, three-dimensional disks 
that have sizable spiral arm density contrasts but nevertheless are stable
to quasi-axisymmetric self-gravitating perturbations.  When a system
is magnetized and has weak net shear, however, it is expected that
perturbations should be able to grow via MJI.

Figure \ref{evol3D} shows that the density 
in our magnetized models ME3d1 and MS3d1 indeed undergoes exponential
growth, with the simulations eventually producing gravitationally bound
condensations.  As we will discuss below, typical wavelengths of the most 
unstable modes in models ME3d1 and MS3d1 are very similar to that of 
the MJI in two-dimensional, thick disks; these wavelengths are much larger than
those of the wiggle instability in two-dimensional models 
(see Tables \ref{tbl-1} and \ref{tbl-2})\footnote{
As Table \ref{tbl-2} indicates, the separation of spurs in our three-dimensional models
depends on the numerical resolution.  Models ME3d2 and MS3d2
with $N_y=128$ form 3 spurs each, while the respective higher resolution 
model with $N_y=256$ forms 4 spurs. 
Based on the results of two-dimensional models (see Table \ref{tbl-1}), which show the 
same number of spurs when $N_y=256$ or 512, 
we expect (but have not proved) that our
three-dimensional models are converged.}.
In addition, spur morphologies and the physical properties of bound clouds 
that form from fragmentation of the spurs in three-dimensional models are similar to those 
due to MJI in two-dimensional thick-disk models.
This supports the case that structure forms  in our three-dimensional models 
as a result of MJI. 
Since gravity is a long-range force and insensitive to density structure
at small scales,
MJI (unlike the wiggle instability) can grow in spite of the
unsteady flows of the background XZ spiral shock. The vertical 
velocity shear and flapping may delay the onset of bound cloud formation 
to some extent, but it is MJI that 
drives three-dimensional systems with magnetic fields into eventual runaway.

As in two-dimensional models, the background flow kinematics 
in our three-dimensional models sculpt the growing 
perturbations into spurs that branch out nearly
perpendicularly from the main spiral arm and become trailing in
interarm regions.
Figure \ref{colb10} displays surface density snapshots at $t/\torb=5.6$, 6.0, 
and 6.3 of model MS3d1 with $\beta_0=10$,
while Figure \ref{colb1} is for model ME3d1 with $\beta_0=1$
at $t/\torb=3.0$, 3.1, and 3.2.
The projected spur morphologies in Figure \ref{colb10}$a$ 
are quite similar to those of model ME2d1 seen in Figure \ref{MHD1pat}$a$,
although the non-stationary nature of XZ spiral shocks tends to wash out
the structure in three-dimensional model ME3d1 compared to the two-dimensional model.
The three-dimensional magnetized models all 
form four spurs, separated by $\lambda_y=L_x/2 = 1.6$ kpc, 
exactly the same as in model ME2d1. 
Thus, three-dimensional effects do not 
significantly alter the development of MJI.  This also implies that 
two-dimensional models with  thick-disk gravity represent an accurate and efficient 
means to explore the MJI without the computational expense of an independent
vertical degree of freedom.

To help visualize the spur structures in three dimensions, we show in Figure 
\ref{spur_vol2}$a$ a perspective volume rendering of an isodensity 
surface and a few selected magnetic field lines 
associated with the region 
marked by the rectangle in Figure \ref{colb10}{\it a} 
(from model MS3d1 at $t/\torb=5.6$ ). 
Shown also in Figure \ref{spur_vol2}$a$ are 
the density slices in logarithmic color scale and the velocity vectors 
at the midplane ($z=0$) as well as at the $y=0.45L_y$ plane;
those at the left edge ($x=0.5L_x$) of the box 
are displayed in Figure \ref{spur_vol2}$b$.
Clearly, the dense part of the spur sticks out perpendicularly from the 
main spiral arm.  The spur has an aspect ratio of $1:0.35:0.16$ in 
($x$, $y$, $z$) and a total mass of $\sim 7\times 10^6\Msun$; 
a similar amount of gas is in the part of the 
spiral arm shown in Figure \ref{spur_vol2}$a$.
Magnetic fields are overall parallel to the spiral
arm, although the variation in the expanding velocity field off the arm 
causes some degree of radial excursions 
of magnetic fields.  This results in magnetic pressure
twice as large in the spur compared to inter-spur regions.
As Figure \ref{spur_vol2}$b$ shows,  magnetic field lines in the 
azimuthal-vertical plane are not strongly bent inward at high $z$,
indicating that  material is accumulated 
mainly along the (azimuthal) $y$-direction.

As spurs develop further, they concentrate  sufficiently to 
trigger gravitational fragmentation, yielding a few bound condensations.
As Figures \ref{colb10} and \ref{colb1} show, fragmentation occurs 
within spiral arms as well as in interarm regions;
fragments forming inside the spiral arms linger  there, whereas 
those forming off the spiral arms are  carried into
the interarm regions and follow galactic rotation.
At the end of the runs, model MS3d1 forms 4 bound clouds 
with an average mass $M\sim1.2\times 10^7\Msun$, which is 
respectively 
about 6 and 1.2 times larger than the local, thin-disk and thick-disk 
Jeans masses at the arm density peak.\footnote{When the 
thick-disk gravity (eq.\ [\ref{kernel}]) is used, one can
show that the Jeans length in a disk with surface density $\Sigma$ and
scale height $H$ becomes $\lambda^{\rm thick}_{\rm J}=
(2\pi \cs^2H/G\Sigma)^{1/2}= (H \lambda_{\rm J})^{1/2}$ when
$2\pi H/\lambda_{\rm J} \gg 1$.  The thick-disk Jeans mass applied to
the peak arm density is then $M_{\rm J,sp}^{\rm thick}=
\Sigma_{\rm sp} (\lambda^{\rm thick}_{\rm J})^2 =2\pi\cs^2H_{\rm sp}/G$,
where $H_{\rm sp}$ is the scale height of gas at the spiral arm peak.}  
These masses are defined by 
\begin{equation}
M_{\rm J,sp}^{\rm thin} = \frac{c_s^4}{G^2\Sigma_{\rm sp}}
= 2.0\times 10^6\Msun
\left(\frac{c_s}{7\kms}\right)^4
\left(\frac{\Sigma_{\rm sp}}{66 \Msun}\right)^{-1},
\end{equation}
and
\begin{equation}
M_{\rm J,sp}^{\rm thick} = \frac{2\pi c_s^2 H_{\rm sp}}{G}
= 1.0\times 10^7\Msun
\left(\frac{c_s}{7\kms}\right)^2
\left(\frac{H_{\rm sp}}{120\; {\rm pc}}\right),
\end{equation}
respectively.
On the other hand, model ME3d1 produces 3 bound clouds, one fewer
than in the corresponding two-dimensional model ME2d1 with equal magnetic field strength.
This is because model ME3d1 with $\beta_0=1$
suffers from strong XZ flapping  of the spiral shock that tends to
work against gravitational mass accumulation.
Bound clouds that form in model ME3d1 have an average mass 
$M\sim3.0\times 10^7\Msun$, comparable to the result from model ME2d1, which is
about 12 times the corresponding thin-disk Jeans mass at the density peak.
Roughly speaking, therefore, MJI  in three-dimensional, magnetized spiral 
shocks turns $\sim20-30\%$ of the total gas into  dense bound clouds in a 
given epoch of gravitational condensation.

Note that typical masses of bound clouds that form in the current 
three-dimensional models are very similar to those in two-dimensional 
thick-disk spiral arm models discussed in \S3.1, as well as 
those in other three-dimensional models without spiral arms 
(Paper II; Paper III). 
In addition, bound clouds from different models share similar magnetic and 
rotational properties.  For example, self-gravitating clouds in our 
three-dimensional models are all magnetically supercritical (by a factor 10 
or more) with an average 
mass-to-flux ratio of $M/\Phi_B\sim2.4 G^{-1/2}$ for model MS3d1 and 
$\sim 1.1 G^{-1/2}$ for model ME3d1.  (Mass-to-flux of $<0.16
G^{-1/2}$ would be subcritical.)
As in two-dimensional cases, these bound clouds in three dimensions are less 
supercritical than the initial simulation boxes, which have 
$M/\Phi_B=7.0 G^{-1/2}$ for model MS3d1 and $2.4 G^{-1/2}$ for model
ME3d1, consistent with the result of \citet{vaz05a}.
These results are remarkably close to that found 
in two-dimensional model ME2d1 in \S3.1 and those resulting from MJI (Paper II)
or via swing amplification (Paper III) in three-dimensional disks without
spiral features.

Because of efficient magnetic torques exerted by field lines linking a
cloud with the surrounding medium (e.g., \citealt{gil74,mou79}),
bound clouds in models MS3d1 and ME3d1 lose a substantial amount of 
angular momentum and thus rotate relatively slowly. 
The mean specific angular momentum of bound clouds at the end of the runs is
$J_z\sim 0.2J_{\rm gal}$ for both models MS3d1 and ME3d1.
Here, $\mathbf{J}=\int \rho(\mathbf{r}\times \mathbf{v})d^3r/
\int \rho d^3r$ with the position $\mathbf{r}$ and velocity $\mathbf{v}$
are measured relative to the cloud center, and
$J_{\rm gal}=\Omega_0(M/\Sigma_0)^2/12$ is the specific
angular momentum contained in a square patch of the galactic ISM.  A 
cloud formed from that patch would preserve the value $J_z=J_{\rm gal}$
 if no angular momentum were lost
during its formation (see Paper III).
Similar specific angular momenta of clouds have been obtained in our 
other three-dimensional magnetized simulations  (Paper II, Paper III),
where we also showed that condensations in unmagnetized models do not 
lose angular momentum as they evolve, while those in magnetized models do.

Finally, we remark on the effects of \citet{par66} instability on the 
formation of spiral-arm substructure in our three-dimensional models. 
The spiral potential compresses the density and magnetic fields,
decreasing the average $\beta$ value inside the spiral shock 
from 10 to 2.5 for model MS3d1 and from 1 to 0.4 for model ME3d1. 
Taking $\beta\sim1$ and the radial wavelength $\lambda_x$ equal to the arm 
width, and neglecting the effect of cosmic ray pressure, the local dispersion 
relation of the Parker instability in a rotating disk presented by 
\citet{shu74} yields $k_y\approx0.5/H$ for the most unstable Parker mode
(this is insensitive to the specific choices of $\beta$ and $\lambda_x$). 
The corresponding wavelength is $\lambda_y =2\pi /k_y = 4 \pi H 
\approx 1.5$ kpc $\approx L_y/4$.  This agrees with the separations identified
for spurs in models MS3d1 and ME3d1.  However, because this wavelength
also coincides with the scale predicted to have the largest growth from
MJI (see \S 3.1 and Fig. \ref{linear}), it is not immediately obvious
whether Parker instability is important in prompting this growth.  

\begin{figure}
\epsscale{1.0}
\plotone{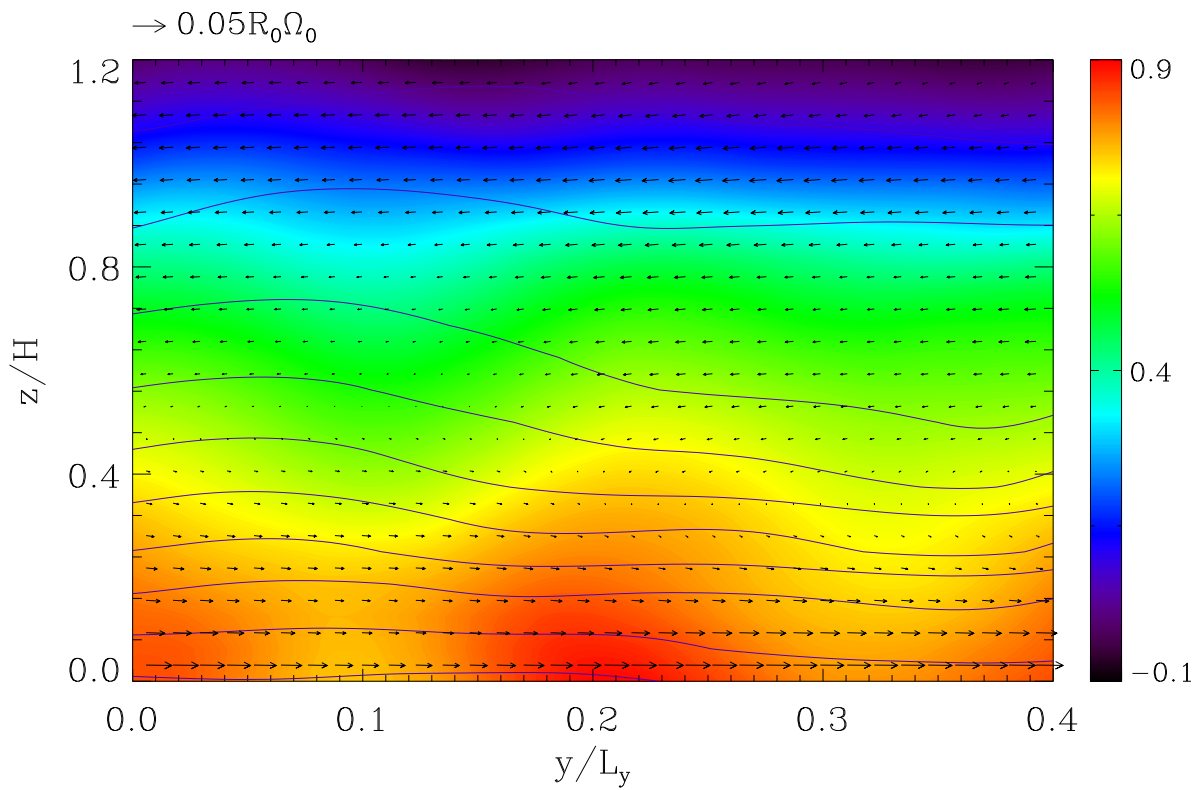}
\caption{YZ slice in the upper half plane at $x/L_x=0.05$ of model MS3d1 at $t/\torb=4.62$ when
spurs are about to emerge.  The density field ($\log(\rho/\rho_0)$ in
color scale) , 
perturbed velocity vectors, and magnetic field lines are drawn.
Strong vertical shear of the azimuthal velocity is apparent.
The characteristics of the Parker instability such as 
the correlation between density and magnetic fields as well as the sinusoidal 
variations in the vertical velocity are not evident.
\label{Parker}}
\end{figure}

To investigate this further, we have examined the detailed
distributions of magnetic fields and velocities during the period when
overdense condensations begin to emerge within the arms.  Figure \ref{Parker}
shows in logarithmic color scale the density in a YZ slice at $x=0.05 L_x$, 
overlaid with a vector
field of perturbed velocity (relative to the mean value in the slice),
and magnetic field lines.  Evidently, there is no sign of the characteristic
correlation of magnetic valleys/hills with overdense/underdense regions that
Parker modes would produce, nor is there any sign of sinusoidal variations
in the vertical velocities with azimuth.  In part, the ability of Parker modes
to grow may be suppressed by the strong vertical gradients in $v_x$ and 
$v_y$ (as seen in Fig.~\ref{Parker} and also Fig. \ref{vshear}).  
Given the curved
nature of the spiral shock front in the XZ plane, high-altitude regions 
can have large horizontal velocities with respect to the midplane gas
below them, such that magnetic fields that begin to 
buckle vertically cannot maintain the shape required for Parker modes to 
develop. 

Thus, we conclude that magnetic buoyancy effects are probably not of
major importance in GMC formation within spiral arms.  Even
without the suppression of Parker modes by vertical shear that we see
here, previous work has shown that Parker instability 
{\it alone} is unable to form high-density clouds in regions away from arms 
(e.g., \citealt{kimj00,san00}; Paper II), 
a consequence of it being a self-limiting
 process stabilized
by tension forces from bent field lines (e.g., \citealt{mou74}).
While Parker instability may help seed structure in early stages
of gravitational instabilities, and may be essential in removing excess 
magnetic 
flux from the disk, it does not appear crucial for massive cloud
formation and subsequent star formation.

\subsubsection{ Fragmentation in Strongly Unstable Arms}\label{AxiU}

We briefly discuss evolution of unmagnetized model HU3d1, with
$Q_0=1.5$, $F=5\%$, and $\beta_0=\infty$.  In this model, 
$Q_0$ is small enough to induce quasi-axisymmetric
gravitational instability of a spiral shock.  
As Figure \ref{evol3D} shows, the maximum density in model HU3d1 at 
$t/\torb=1.5$ is larger by a factor of 3 than that in stable model H3d1 
with $Q_0=1.8$.  
Soon after the spiral perturbation attains its full strength, the system
is dominated by the exponential growth of quasi-axisymmetric modes in the 
spiral shock.
At around $t/\torb=2.2$, the collapsing post-shock layer reaches 
sufficient density that it begins to undergo non-axisymmetric 
gravitational fragmentation.  This results in the formation of 
13 bound condensations aligned along the shock front,  
with mass $\sim1\times 10^7\Msun$ each.
Note that while magnetized models MS3d1 and ME3d1 first develop spur
structures that subsequently fragment into bound clouds in both arm
and interarm regions, the bound clouds in unmagnetized model HU3d1 
do not require prior formation of spurs.  These clouds form directly
in the highest-density ridge
of the spiral shock. 
The spacing of bound clouds in model HU3d1 is $\lambda/\lambda_{\rm J,0}
\sim 0.5$, where $\lambda_{\rm J,0}=\cs^2/(G\Sigma_0)$ is the Jeans length
in the initial featureless disk.  This  
can be compared with $\lambda/\lambda_{\rm J,0}\sim1.7$
for the spur separation in magnetized model MS3d1 with the same $Q_0$ and 
$F$ parameters as model HU3d1 (see Table \ref{tbl-2}).\footnote{Since in 
axisymmetry there is no stable shock for the
parameters of model HU3d1, we cannot measure cloud spacings in terms of
$\lambda_{\rm sp}$ (which is undefined).}
This demonstrates that the presence of magnetic fields makes
significant differences in the dynamical evolution of gas in spiral arms.

\section{Conclusions}

\subsection{Summary}

A key unsolved problem in the dynamics and evolution of spiral
galaxies lies in understanding the origin of 
spiral-arm substructure, including gaseous spurs, giant molecular
clouds, and complexes of giant \ion{H}{2} regions.  Various 
observational and theoretical arguments increasingly support the notion
that this arm substructure may originate from 
gravitational instability of diffuse gas within spiral arms.  
In Paper I, we studied dynamical evolution of a local segment of
a magnetized spiral arm in a razor-thin model of a disk. We showed that 
magneto-Jeans instability (MJI) initiated within spiral arms naturally
yields gaseous spurs extending into the interarm region.
Paper I further showed that these spurs fragment into bound 
condensations that could potentially evolve into arm and interarm
\ion{H}{2} regions.

The thin-disk models of Paper I, however, tend to overestimate midplane 
self-gravity, which favors small-scale MJI modes.  The models of Paper I 
were also unable to capture potential dynamical 
consequences of the Parker instability and other three-dimensional
instabilities that rely critically upon the vertical dimension.  
Here, we have extended Paper I to consider these important effects.
We consider two sets of numerical models: 
two-dimensional disks in which a thick-disk gravitational kernel 
(eq.\ [\ref{kernel}])
approximately treats the geometric dilution of self-gravity,
and full three-dimensional disks in which all fluid variables are allowed to 
vary with the vertical coordinate.  
As in Paper I, both models adopt an isothermal equation of state.
Our main objectives were to explore how the properties of spurs 
and bound condensations produced by MJI in vertically-extended disks 
differ from the models of Paper I, and to examine the effectiveness of
three-dimensional dynamical instabilities other than MJI in forming
spiral-arm substructures.
The following summarizes the main results of the present work:

1.  In our new models with sufficiently-magnetized conditions, spiral
shocks give rise to gaseous spurs and bound clouds in a manner similar
to the models of Paper I.  This is true either for two-dimensional
``thick disk'' or three-dimensional models.  The spur structures
themselves are slightly more trailing than those in razor-thin disks.
Dilution of self-gravity due to finite disk thickness is significant, causing
the typical separation of spurs $\lambda_y$ (along the arm) to be about
10 times the Jeans length $\lambda_{\rm J,sp}=c_s^2/(G \Sigma_{\rm sp})$
at the spiral arm peak, which is 3 -- 5 times larger than the prediction of
$\lambda_y/\lambda_{\rm J,sp}$ from thin-disk models.
The agreement between three-dimensional and two-dimensional ``thick disk'' 
models implies that, in this strong-$B$-field case, the mechanism behind 
spur and clump formation is the MJI.  
Reduced gravity in thick disks also causes the bound condensations that form 
when spurs fragment to be more massive than in thin-disk models.  The average 
mass of these clouds is $\sim(1-3)\times 10^7\Msun$ corresponding to 6 -- 12 
times the thin-disk Jeans mass at the arm peak, and comparable to the 
thick-disk Jeans mass at the arm peak.
The clump masses are comparable to the thin-disk Jeans mass at the mean
unperturbed surface density of the disk.
These bound clouds are magnetically supercritical with the mean mass-to-flux 
ratio of $\sim (1-3)G^{-1/2}$, and undergo significant loss of angular momentum
($80\%$ of the initial galactic value) via magnetic braking.

2.  Before gravitational instability sets in, our three-dimensional
models exhibit time-dependent behavior in their spiral shock
structure that is quite unlike the rapid approach to steady state that
characterizes two-dimensional models.  The three-dimensional
distributions can be averaged over the $y$ direction (parallel to the
spiral arm) to yield an XZ shock profile.  In these XZ profiles, the
shock front generally moves back and forth relative to its mean
position (in quasi-radial coordinate $x$).  The flapping period of the
XZ spiral shock in an unmagnetized model is $\sim 0.3\torb$,
comparable to the epicyclic period at the arm peak.  The amplitude of
the flapping motion tends to be larger at large $z$ in a given model,
with radial excursions of $\sim0.06L_x$ at $|z|/H_0=1$ in the
$\beta_0=1$ model.  This amplitude is a factor 3 or 2 times larger
than in models with $\beta_0=10$ or unmagnetized ($\beta_0=\infty$)
models, respectively.  This flapping and other nonsteady motions do
not seem to strongly affect development of gravitational
instabilities, however, probably because they depend on a mean density
enhancement over large scales that need not be strictly coherent for growth. 

3. It is informative to consider the time average of the XZ shock
profiles.  The shock front in this mean profile is in general curved
in the $x$-$z$ plane.  In these XZ profiles, spiral arm regions are
generally thinner than interarm regions; the ratio of scale heights in
arm to interarm regions is about 0.5 for models with $\beta_0=10$ or
$\infty$ and $\sim 0.8$ for $\beta_0=1$ models.  The surface density
distributions of the time-averaged profiles are similar to those of
one-dimensional steady spiral shocks resulting from two-dimensional
thick-disk-gravity models averaged over $y$.  This suggests that the
thick-disk gravitational kernel of equation (\ref{kernel}) provides a
good approximation for the gravitational potential near the midplane
of a three-dimensional disk.  It also shows why, due to the right {\it
  average} conditions, gravitational instabilities grow,
despite significant stochasticity in the flow. 

4. We find that weakly magnetized or unmagnetized two-dimensional
models are unstable to the ``wiggle instability'' described by
\citet{wad04} (based on non-self-gravitating models with strong
shocks).  \citet{wad04} advocated Kelvin-Helmholtz instabilities as an
explanation for the wiggle instability, and argued that spiral arm
spurs could arise from this mechanism.  Indeed, we find the magnetic
field topology and the generation of vorticity near the shock front in
our two-dimensional unmagnetized or weakly magnetized models are
suggestive of Kelvin-Helmholtz instabilities in shear layers.  We also
find that mergers of vortical clumps in these two-dimensional models
can eventually produce self-gravitating clouds with mass $\sim 1\times
10^7\Msun$, although the spur structures are much less prominent than
in our two-dimensional models with stronger magnetic fields.

Most importantly, however, we find that the vorticity-generating wiggle
instability is absent in full three-dimensional models of all magnetic
field strengths.  It appears that radial flapping motions of the XZ
shock front, combined with strong vertical shear of horizontal
velocities, quickly disrupt any coherent vortical structures that
would otherwise grow.  We therefore conclude that the wiggle
instability is an artifact of two-dimensional models within a certain
parameter range, and is unlikely to play an important role in forming
spiral-arm substructures in real spiral galaxies.

5.  While the Parker instability has long been invoked as a primary
mechanism for the formation of giant molecular clouds inside spiral
arms, our three-dimensional models do not show any noticeable evidence
of developing Parker modes.  There is no indication of sinusoidal
vertical velocities, or correlation of magnetic hills/valleys with
over/under dense regions.  Like the wiggle instability, the Parker
instability appears to be suppressed by strong vertical shear of
inplane velocities.

6.  In addition to MJI, two other mechanisms capable of forming
gravitationally bound clouds in disk galaxies are swing amplification
and collapse of spiral arms parallel to the shock front (followed by
fragmentation).  The simulation outcomes of our unmagnetized,
three-dimensional models with spiral arms show, however, that the
growth of nonaxisymmetric disturbances via swing amplification is very
small when $F=5\%$ and $Q_0=1.8$.  This is consistent with the results
of Paper I, in which we showed that swing amplification in thin disks
is efficient only if the background spiral perturbation does not
exceed 1\%.  With reduced self-gravity, swing amplification in
three-dimensional spiral arms becomes even more inefficient. Thus, the
swing mechanism would appear to apply either in interarm regions or in
galaxies without strong spiral arms.

On the other hand, our unmagnetized model with $F=5\%$ and $Q_0=1.5$ was
sufficiently unstable that quasi-axisymmetric growth developed as soon as
the external potential reached full strength.  Fragmentation into
closely-spaced bound clouds then occurred.  Conceivably, this kind of
cloud formation process could be important in galaxies that are
intrinsically quite close to instability, and experience a tidal
encounter that tips them ``over the edge''.

\subsection{Discussion}

The summary above makes clear that our new three-dimensional models
uphold the principal conclusions of Paper I -- namely, that
gravitational instability in the gas component of spiral arms (1) is
able to form bound clouds with properties similar to the most massive
GMCs, and (2) can lead to the development of interarm extensions
similar to features that have been described in the observational
literature as spiral arm spurs and feathers.

\begin{figure}
\epsscale{1.0}
\plotone{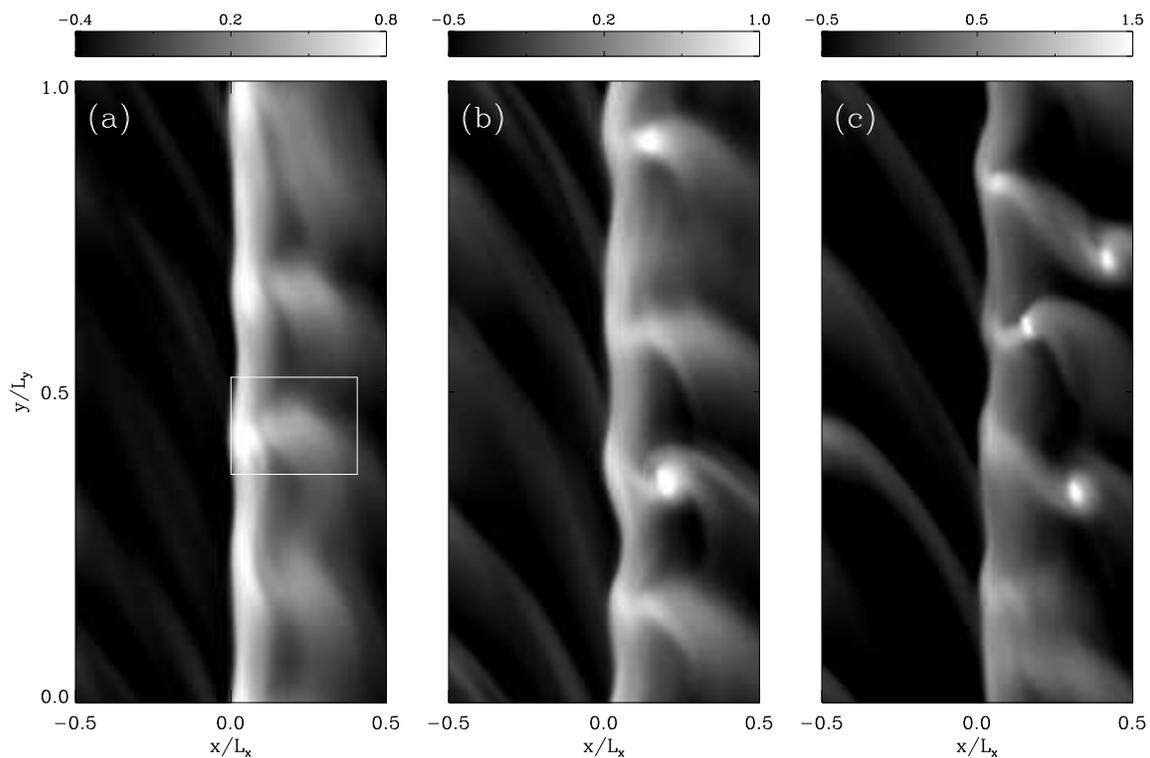}
\caption{Snapshots of vertically-integrated surface density 
($\log(\Sigma/\Sigma_0)$ in gray scale) of $\beta_0=10$ model MS3d1
at ($a$) $t/\torb=5.6$,  ($b$) $t/\torb=6.0$, 
and ($c$) $t/\torb=6.3$.
For fiducial parameters,
the simulation box in the $x$-$y$ plane has a size of $L_x=L_y/2=3.14$ kpc.
The rectangle in ($a$) indicates the sector viewed as 
a three-dimensional visualization in Fig.\ \ref{spur_vol2}. 
\label{colb10}}
\end{figure}

\begin{figure}
\epsscale{1.0}
\plotone{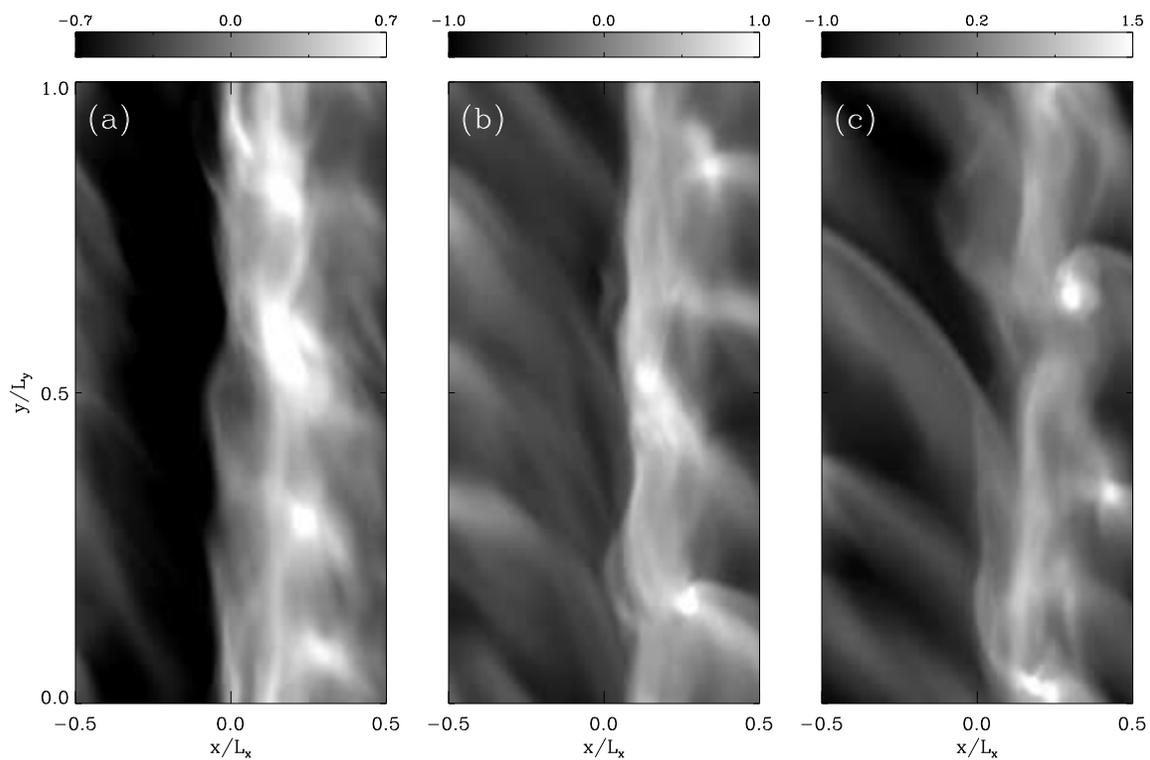}
\caption{Snapshots of surface density ($\log(\Sigma/\Sigma_0)$ in gray 
scale) of $\beta_0=1$ model ME3d1 at ($a$) $t/\torb=3.0$, 
($b$) $t/\torb=3.1$, and ($c$) $t/\torb=3.2$.  
For fiducial parameters,
the simulation box has a size of $L_x=L_y/2=3.14$ kpc.
Due to the strong radial flapping motions of gas near the
spiral shock, spurs in model ME3d1 are less conspicuous than in
model MS3d1.
\label{colb1}}
\end{figure}

Our models show that gaseous spurs may continue to stand out against
the interarm background for a long time after clouds form, without
being dispersed (see Figures \ref{colb10} and \ref{colb1}).  Since
interarm regions are characterized by low gas density and strong
shear, they are normally thought of as being hostile for gas to
condense gravitationally.  Our model simulations predict, however,
that even interarm regions with low {\it mean} gas density can be
abundant with kpc- and larger-scale substructure, arranged in trailing
gaseous spurs that may host \ion{H}{2} regions.  The material that
makes up these interarm spurs consists of parcels of gas that were the
center of attraction during the most recent spiral arm transit.  The
concentration of interarm gas into secondary spurs may be crucial in
enabling clouds, and hence stars and \ion{H}{2} regions, to form at
large distances from the gas-dust ridges that mark the primary loci of
spiral arms.  Because of this, azimuthal offsets between
primary dust lanes and maxima in H$\alpha$ emission may not give
meaningful measures of star formation ``lag'' timescales (cf. \cite{mou05}). 

While the bound clouds produced in our models continue collapsing in a
runaway fashion, our numerical resolution is inadequate to follow
their evolution beyond the point where the central cloud density has
grown by more than two orders of magnitude.  This is because the local
Jeans length must be resolved with at least a few spatial zones for
accurate evolution; unphysical fragmentation may otherwise occur
\citep{tru97,tru98}.  Although the future of the massive, bound clouds
we identify cannot be known at this point, we speculate that their
internal turbulent dynamics (unresolved in the present models) would
lead to fragmentation into GMCs, some portions of which would undergo
clustered star formation.  The eventual result would be giant
\ion{H}{2} regions in both spiral arm and interarm regions.

The masses, magnetic fluxes, and angular momenta of the clouds formed
via gravitational instability in the present three-dimensional models
are quite similar to the results of Papers II and III, in spite of
differences in the specific mechanisms involved in each case.  In
Paper II, there was no spiral structure present, while in Paper III
there was strong turbulence associated with the magnetorotational
instability.  Thus, we conclude that it would be difficult to discern
the detailed route to cloud formation by observing the end result; not
enough information is retained.  On the other hand, the similarity of
these bulk cloud properties from models with rather different
particulars indicates a robustness of conclusions for the outcome of
self-gravitating cloud formation in all kinds of disk galaxies.

Our finding that gas in spiral arms is thinner 
than in interarm regions appears to be
consistent with \citet{nak03} who reported that the 
scale height of the \ion{H}{1} layer in our Galaxy systematically decreases 
with increasing midplane density. 
The fact that the Milky Way molecular gas, which predominantly
 traces spiral arms,
has about half the scale height of \ion{H}{1} gas 
\citep{mal94}
also supports our result.  \citet{sta05a} furthermore find that massive
CO GMCs, which are preferentially associated with spiral arms in the 
Milky Way \citep{sol85,sol89}, have a smaller scale height than low-mass 
molecular clouds.
We note however that 
more quantitative comparison would require consideration of
ISM turbulence and other processes that may 
vary spatially, and significantly affect the gas scale height.

In the present models, we find that 20-30\% of the total gas has 
collected into massive bound clumps by the end of the simulation, in
those cases where gravitational instability occurs.  We have found
similar, or slightly smaller fractions in the uniform-shear ($q_0=1$) 
three-dimensional models of Paper II and III (the range is $\sim 15-20\%$).
This fraction can be thought of as the efficiency, per GMC formation
timescale, of converting diffuse gas to gas capable of forming stars.  In
the present models, the timescale for gravitational instability to run
away is 2-6 orbits, for a range of values of $Q_0$, $F$, and $\beta_0$.
Combining these raw numbers would imply an effective GMC formation 
timescale from diffuse gas of $\sim 10t_{\rm orb}$.  However, we believe
that in fact this somewhat overestimates the true timescale for the
average parcel of diffuse gas to be incorporated in a bound cloud.
While the conversion efficiency per GMC formation timescale may be
realistic, the time to reach runaway is probably longer than it would
be in a real system, because our initial perturbation amplitudes are likely 
low compared to realistic values.  An important direction for future
study will be to measure cloud formation rates in simulations where cloud
destruction is also incorporated; we plan to pursue studies of this kind.
Models of this kind will provide more realistic background conditions
from which gravitational instabilities can develop.

While the real ISM has a multi-phase structure due to the particular
heating and cooling processes that are relevant, we have adopted a
much simpler isothermal approximation throughout this paper.  The
inclusion of gaseous cooling that leads to dense cloud formation at
small spatial scales (e.g., \citealt{koy02,pio04,pio05,aud05,
  heit05,vaz05b}) might indirectly affect formation timescales, mean
separations, and masses of the GMCs and other structures that result
from self-gravitating instabilities in spiral arms.  It is often
assumed that the total velocity dispersion in the warm/cold medium can
be treated as an effective sound speed; whether this treatment is
adequate will be addressed directly in future work.  Our
three-dimensional isothermal simulations show that the wiggle
instability is stabilized by strong vertical shear and non-steady
flows associated with spiral shocks, but the situation could
conceivably be different if thermal instability and multi-phase gas
structure are considered.  This outstanding issue will direct our
future research as well.

Finally, we remark that, while spiral structure clearly can help
gravitational instability and prompt star formation, it may also
contribute to limiting these processes.  As we have discussed, spiral
shocks (especially when strongly magnetized), lead to time dependent
motions that may be important in feeding turbulence at scales $\ll H$.
The present models are isothermal at $T\sim 8,000$K, so the sound speed
$7\kms$ exceeds turbulent velocity amplitudes.  However, in the real
ISM there is significant cold gas for which the random turbulent
motions far exceed the sound speed $\sim 1\kms$.  The turbulence in
the cold ISM component may be key to regulating self-gravitating
instabilities.  In a future publication, we will discuss results of an
investigation focusing on the details of turbulent driving by spiral
shocks.

\acknowledgments

We gratefully acknowledge helpful discussions with S.~S.~Hong, J.~Kim, 
D.~Ryu, J.~Stone, and S.~Vogel.
We are also grateful to the referee, E.\ V\'azquez-Semadeni, for 
a thorough and insightful report. 
W.-T.~Kim was supported by a New Faculty Grant at the SNU.  The work of 
E.C. Ostriker was supported by NSF grants AST 0205972 and 0507315.
This work was also supported in part by NASA grant NNG 05GG43G.
Numerical simulations were performed on the O2000 system at the NCSA, 
the Linux cluster at KASI (Korea Astronomy and Space Science Institute) 
built with funding from KASI and ARCSEC, and the IBM p690 at the KISTI.

\clearpage
\begin{deluxetable}{ccccccccccc}
\tabletypesize{\footnotesize}
\tablecaption{Parameters of Two-Dimensional Models with Thick-Disk Gravity
\label{tbl-1}}
\tablewidth{0pt}
\tablehead{
\colhead{\begin{tabular}{c}Model\tablenotemark{a} \\ (1) \end{tabular} } & 
\colhead{\begin{tabular}{c}$Q_0$           \\ (2) \end{tabular} } &   
\colhead{\begin{tabular}{c}$\beta_0$       \\ (3) \end{tabular} } & 
\colhead{\begin{tabular}{c}$F$             \\ (4) \end{tabular} } & 
\colhead{\begin{tabular}{c}Resolution      \\ (5) \end{tabular} } & 
\colhead{\begin{tabular}{c}$\Sigma_{\rm sp}/\Sigma_0$ 
                                           \\ (6) \end{tabular} } & 
\colhead{\begin{tabular}{c}$Q_{\rm sp}$    \\ (7) \end{tabular} } & 
\colhead{\begin{tabular}{c}$W/L_x$
                                           \\ (8) \end{tabular} } & 
\colhead{\begin{tabular}{c}$\lambda_y/L_x$ 
                                           \\ (9) \end{tabular} } & 
\colhead{\begin{tabular}{c}$K_{y}$
                                           \\(10) \end{tabular} } 
}
\startdata
ME2d1        &  1.2            & 1          &
10\%         & $256\times512$  & 4.20       &  
0.59         & 0.15            & 0.50       & 0.11
\\
ME2d2        &  1.2            & 1          &
10\%         & $128\times256$  & 4.14       &  
0.59         &  0.15           & 0.50       & 0.11
\\
ME2d3        &  1.2            & 1          &
10\%         & $128\times128$  & 4.14       & 
0.59         &  0.15           & 0.67       & 0.08
\\
\\
MS2d1           &  1.5            & 10         &
5\%          & $256\times512$  & 5.96       &  
0.61         & 0.062           & 0.22       & 0.22
\\
MS2d2        &  1.5            & 10         &
5\%          & $128\times256$  & 5.53       &  
0.64         &  0.069          & 0.22       & 0.23
\\
MS2d3           &  1.5            & 10         &
5\%          & $128\times128$  & 5.53       &  
0.64         &  0.069          & 0.33       & 0.16
\\
\\
H2d1         &  1.8            & $\infty$   &
5\%          & $256\times512$  & 6.75       &  
0.69         &  0.044          & 0.16       & 0.32
\\
H2d2         &  1.8            & $\infty$   &
5\%          & $128\times256$  & 6.33       &  
0.72         &  0.049          & 0.20       & 0.27
\\
H2d3         &  1.8            & $\infty$   &
5\%          & $128\times128$  & 6.33       &  
0.72         &  0.049          & 0.31       & 0.18
\\
\enddata

\tablenotetext{a}{The prefixes ME, MS, and H stand for the
magnetized models with equipartition ($\beta_0=1$) and sub-equipartition 
($\beta_0=10$) field strengths, and hydrodynamic models.}
\end{deluxetable} 

\clearpage
\begin{deluxetable}{ccccccccccc}
\tabletypesize{\footnotesize}
\tablecaption{Parameters of Three-Dimensional Simulations
\label{tbl-2}}
\tablewidth{0pt}
\tablehead{
\colhead{\begin{tabular}{c}Model           \\ (1) \end{tabular} } &
\colhead{\begin{tabular}{c}$Q_0$           \\ (2) \end{tabular} } &
\colhead{\begin{tabular}{c}$\beta_0$       \\ (3) \end{tabular} } &
\colhead{\begin{tabular}{c}$F$             \\ (4) \end{tabular} } &
\colhead{\begin{tabular}{c}Resolution      \\ (5) \end{tabular} } &
\colhead{\begin{tabular}{c}$\Sigma_{\rm sp}/\Sigma_0$
                                           \\ (6) \end{tabular} } &
\colhead{\begin{tabular}{c}$Q_{\rm sp}$    \\ (7) \end{tabular} } &
\colhead{\begin{tabular}{c}$H_{\rm arm}/H_0$
                                           \\ (8) \end{tabular} } &
\colhead{\begin{tabular}{c}$\lambda_y/L_x$
                                           \\ (9) \end{tabular} } &
\colhead{\begin{tabular}{c}$K_{y}$
                                           \\(10) \end{tabular} }
}
\startdata
ME3d1        &  1.2                     & 1          &
10\%         & $128\times256\times128$  & 3.32       &
0.66         & 0.93                     & 0.50       & 0.14
\\
ME3d2        &  1.2                     & 1          &
10\%         & $128\times128\times128$  & 3.30       & 
0.66         & 0.93                     & 0.67       & 0.10
\\
\\
MS3d1        &  1.5                     & 10         &
5\%         & $128\times256\times128$  & 5.31       &
0.65         & 0.71                     & 0.50       & 0.11
\\
MS3d2        &  1.5                     & 10         &
5\%         & $128\times128\times128$  & 5.30       &
0.65         & 0.72                     & 0.67      & 0.08
\\
\\
H3d1        &  1.8                     & $\infty$   &
5\%         & $128\times256\times128$  & 5.06       &
0.80         & 0.64                     & ...       & ... 
\\
H3d2        &  1.8                     & $\infty$    &
5\%         & $128\times128\times128$  & 5.02        &
0.80         & 0.66                     & ...        & ... 
\\
\\
HU3d1\tablenotemark{a}        &  1.5                     & $\infty$   &
5\%          & $128\times256\times128$  & ...                      & 
...                           & ...     & 0.15       & 1.92\tablenotemark{b}
\enddata

\tablenotetext{a}{The prefix HU stands for hydrodynamic model with
a strongly unstable arm.}
\tablenotetext{b}{Relative to the initial background disk without
spiral potential perturbations.}

\end{deluxetable}

\clearpage

\clearpage

\clearpage

\clearpage

\clearpage

\clearpage

\clearpage

\clearpage

\clearpage

\clearpage

\clearpage

\clearpage
\begin{figure}
\epsscale{1.0}
\vspace{-2cm}
\caption{({\it a}) 
Volumetric rendering (seen from above the galactic plane)
of an isodensity surface at $\rho=2.5\rho_0(0)$, together with 
selected magnetic field lines, from a portion of 
model MS3d1 at $t/\torb=5.6$.
The region fits in a projected rectangle marked in Fig.\ \ref{colb10},
with $|z|/H\! <\!1.25$.  
The magnetic field lines in blue
run from $(y,z)=(0.35L_y,0)$ at the lower edge of the box toward the upper
edge, while yellow denotes field lines originating at 
$(y,z)=(0.35L_y,0.56H)$.
The density (($\log(\rho/\rho_0)$ in color scale) and velocity vectors at $z=0$ are 
shown
at the bottom of the box, while those at $y=0.45L_y$
are displayed on the side wall (top of page).
({\it b}) The density ($\log(\rho/\rho_0)$ in color scale), velocity vectors, and 
magnetic field lines from 
the left surface ($x/L_x=0.5$) of the box shown in (a).
In both panels, the velocity vectors are relative to the
center of the box, where the density attains its maximum.
\label{spur_vol2}}
\end{figure}

\clearpage

\end{document}